\documentclass[aps,prd,11pt, tightenlines, twoside, secnumarabic, superscriptaddress, showpacs, preprintnumbers, nofootinbib, notitlepage, onecolumn, fleqn]{revtex4-1}

\pdfoutput=1
\usepackage[english]{babel}
\usepackage{amsmath,amssymb,amsfonts, bm,bbm,slashed}
\usepackage{graphicx}
\usepackage[sort&compress]{natbib}
\usepackage[dvipsnames]{xcolor}
\usepackage[normalem]{ulem}
\usepackage{hyperref}
\usepackage{cleveref}
\definecolor{red}{rgb}{1.0, 0, 0}
\usepackage{hyperref}
\usepackage{enumerate}
\usepackage{epsfig, subfigure}
\usepackage{setspace}
\usepackage{booktabs, tabularx}
\usepackage{units}

\hypersetup{
    pdfnewwindow=true,   
    colorlinks=true,     
    linkcolor=blue,      
    citecolor=blue,      
    filecolor=blue,      
    urlcolor=blue        
}

\allowdisplaybreaks

\bibpunct{[}{]}{,}{n}{}{,}

\providecommand{\abs}[1]{\lvert#1\rvert}

\newcommand{\be}[2]{\left.  #1 \right|_{#2}}              

\newcommand{\vev}[1]{\left\langle  #1 \right\rangle }      

\newcommand{\GeV}{\,\text{GeV}}
\newcommand{\TeV}{\,\text{TeV}}

\begin{document}

\title{Variations on the Vev Flip-Flop: \\Instantaneous Freeze-out and Decaying Dark Matter}

\author{Michael J.\ Baker}
\email{baker@physik.uzh.ch}
\affiliation{Physik-Institut, Universit\"at Z\"urich, 8057 Z\"urich, Switzerland}

\author{Lukas Mittnacht}
\email{lmittnac@students.uni-mainz.de}
\affiliation{PRISMA Cluster of Excellence \& Mainz Institute for Theoretical Physics,
        Johannes Gutenberg University, Staudingerweg 7, 55099 Mainz, Germany}
             
\date{\today}
\preprint{MITP/18-109}
\preprint{ZU-TH-41-18}


\begin{abstract}
In this work we consider a simple model for dark matter 
and identify regions of parameter space where 
the relic abundance is set via kinematic thresholds, 
which open and close due to thermal effects.  
We discuss \emph{instantaneous freeze-out}, where dark matter 
suddenly freezes-out when the channel connecting dark matter to the thermal bath closes, 
and \emph{decaying dark matter}, where dark matter freezes-out while relativistic 
and later decays when a kinematic threshold temporarily opens. 
These mechanisms can occur in the vicinity of a one-step or a 
two-step phase transition. 
In all cases thermal effects provide this dynamic behaviour, while 
ensuring that dark matter remains stable until the present day.  
\end{abstract}

\maketitle

\section{Introduction}
\label{sec:intro}

Understanding the 
nature of dark matter (DM) is one of the most pressing and elusive problems in physics. 
For a long time, the best-motivated candidate has been
the Weakly Interacting Massive Particle (WIMP).  In this picture, DM particles 
are hypothesised to be heavy ($m_\text{DM} \gtrsim 100$~GeV) and to have weak but
non-negligible interactions with Standard Model (SM) particles.  In the
very early universe, these interactions keep the DM and the SM particles 
in thermal and chemical equilibrium, until eventually the expansion of the universe dilutes
the DM density to the extent that DM annihilation into
SM particles ceases, and a relic abundance survives to the present day (a process known as freeze-out). 
However, in spite of a vigorous, multi-pronged search 
programme~\cite{Aaboud:2017phn,Sirunyan:2017hci,Aprile:2018dbl,Ahnen:2016qkx,Abdallah:2016ygi,Aartsen:2017ulx}, 
there is still no evidence for WIMPs, and the 
remaining allowed parameter space has been drastically reduced.

There is currently a large effort studying alternative mechanisms of DM 
production~\cite{Tanabashi:2018oca}.  
In the early universe, the SM particles formed a hot, thermal bath 
and the effects of this plasma can have a dramatic impact on 
particle properties and interactions.
These finite-temperature effects are known to cause important phenomena 
in the development of the universe, such as the electroweak phase transition (EWPT).
Recently there has been interest in understanding 
the impact of finite temperature effects in mechanisms of DM 
production~\cite{Rychkov:2007uq, Strumia:2010aa, Hamaguchi:2011jy, Drewes:2015eoa, Baker:2016xzo, 
Kobakhidze:2017ini,Baker:2017zwx,J.Baker:2018eaq,Hektor:2018esx,Bian:2018mkl}.  

In the present work we examine the mechanism presented in~\cite{Baker:2016xzo,J.Baker:2018eaq}   
where the DM abundance is set not via freeze-out but via a period of DM decay, 
which occurs when a new scalar field temporarily obtains a vacuum expectation value (vev) just before the EWPT. 
We find that the key feature required for this period of decay is the opening 
and closing of kinematic thresholds, due to particle masses obtaining 
temperature dependent corrections.  
In the early universe, bosons receive large finite temperature 
corrections and their masses and vevs should be treated as functions of 
temperature.
In \cref{sec:model}, we propose a simple model which 
retains the key features of the mechanism in~\cite{Baker:2016xzo}. 
With this simple model, we identify further regions of parameter space 
where interesting DM production mechanisms can appear.  
We first discuss the effective potential at 
finite temperatures in \cref{sec:effpot}, 
the thermal bath in \cref{sec:thermal-bath} and 
Boltzmann equations to track the DM abundance in \cref{sec:dmdecay}.  
With this framework, we describe \emph{instantaneous freeze-out} in \cref{sec:instantaneous}, 
where DM freezes-out abruptly when a kinematic channel closes.  
In \cref{sec:one-step-decay} we show that a period of DM decay can set the relic abundance,
dubbed \emph{decaying dark matter},
when a new scalar field obtains a vev in a one-step phase transition,
while in \cref{sec:two-step-decay} we look at a situation 
similar to that described in~\cite{Baker:2016xzo} 
where a two-step phase transition (or a vev flip-flop) can lead to a period of 
decaying dark matter. 
Finally, in \cref{sec:exp-const} we briefly consider the experimental constraints on the different scenarios.

\section{The Model}
\label{sec:model}

\begin{table}
  \centering
  \begin{minipage}{10cm}
    \begin{ruledtabular}
    \begin{tabular}{cccc}
      Field  &      Spin     & $\mathbb{Z}_2$ & mass scale (at $T=0$) \\
      \hline
      $S$    &        0           &      $+1$      &  $ \sim 0.1 \text{ -- } 100\GeV$ \\
      $\chi$ & $\frac{1}{2}$ &      $-1$      &  $ \sim 5\GeV \text{ -- } 5\TeV$ \\
      $\psi$ & $\frac{1}{2}$ &      $-1$      &  $ \sim 5\GeV \text{ -- } 5\TeV$ \\
      \bottomrule
    \end{tabular} 
    \end{ruledtabular}
  \end{minipage}
  \caption{The new particles we introduce in \cref{sec:model} with their respective charges 
    and mass scales.  All new particles are SM gauge singlets.}
  \label{tab:particles}
\end{table}

To demonstrate the effects as clearly as possible, we consider a simple model consisting 
of the SM plus a real scalar, $S$, and 
two dark sector Dirac fermions, $\chi$ and $\psi$, shown in \cref{tab:particles}.  
In what follows, $\chi$ will constitute the dominant component of 
 dark matter.
 These particles will all be gauge singlets, and will have masses 
 in the GeV to TeV range.
 
The Lagrangian for these fields is
\begin{align} 
  \mathcal{L} &\supset
      \frac{1}{2} (\partial_\mu S)(\partial^\mu S)
    + \bar{\psi}(i\slashed{\partial} - \tilde{m}_\psi)\psi 
    + \bar{\chi}(i\slashed{\partial} - m_\chi)\chi
                                            \nonumber\\[0.2cm]
  &\qquad
    - \big[ y_{\chi\psi} \, \bar\psi S \chi + h.c. \big]
    - y_\chi \, \bar\chi S \chi
    - y_\psi \, \bar\psi S \psi
    - V(H, S)\,,
  \label{eq:kin:L}
\intertext{with}
  V(H, S) &=
   - \mu_H^2 H^\dag H
   + \lambda_{H} \, (H^\dag H)^2
   - \frac{1}{2} \mu_S^2 S^2
   + \frac{\lambda_{S}}{4!} S^4
   + \frac{\lambda_{p}}{2} \, S^2 (H^\dag H) \,.
  \label{eq:kin:V} 
\end{align}
In our notation, the SM Higgs potential has $\mu_H \simeq 88\GeV$ and 
$\lambda_H \simeq 0.12$, with $H=(G^+, (h+iG^0)/\sqrt{2})$.
We choose a basis where the CP-even, neutral 
component $h$ obtains the vev.
We see in \cref{eq:kin:V} that the only interaction at dimension-4 between these new particles and the 
Standard Model is the Higgs portal term $(\lambda_p/2) \, S^2 (H^\dag H)$.  
Although in principle there may also be the terms $S^3$ and $S (H^\dag H)$, 
we assume these couplings to be negligibly small.
In  \cref{sec:instantaneous,sec:one-step-decay} we will 
consider $\lambda_p \ll 1$.  
In this regime the scalar fields decouple and we can consider the evolution 
of $S$ alone, simplifying the picture.  
In \cref{sec:two-step-decay} we will consider $\lambda_p \simeq 1$.
For the new Yukawa couplings we will typically 
consider $y_\psi \simeq 1$, $y_{\chi\psi} \sim 10^{-7}$ and 
$y_\chi \simeq 0$.  A large $y_\psi$ means that $\psi$ will remain in 
thermal equilibrium throughout the processes affecting the $\chi$ abundance. 
This small $y_{\chi\psi}$ will however make direct and indirect detection 
of dark matter challenging.
Within this model, none of the operators with small couplings will be 
generated at a significant level via loops.

We will focus on the regions of parameter space where finite temperature 
effects from the scalar sector have a significant impact on the final 
abundance of $\chi$.  
We know from the standard WIMP picture that the observed relic abundance 
is obtained if DM freezes-out when its mass is 20 -- 30 times greater 
than the freeze-out temperature.  
Since we do not have abundances below the equilibrium abundance, this provides a lower limit on $m_\chi$.  
If the DM mass is much greater than the temperature at which 
the abundance is set, then finite temperature effects will be small compared 
to zero temperature effects.  
This tension means that for the mechanisms presented in this paper, 
we will generally focus on the parameter space where   
$m_\chi, \tilde{m}_\psi \simeq 30\, \mu_S$. 
We choose the signs of the Yukawas so that for positive couplings, 
$\vev{S} \neq 0$ will give a positive contribution to the mass.
In \cref{tab:summary} we summarise the different mechanisms and highlight 
the relevant parameter space.

Finally we emphasise that many models can accommodate degenerate or nearly-degenerate vector-like fermions, while the hierarchy between the masses of the new fermions and the new scalar may be expected since their mass scales are not directly connected.  Similarly, while the strong hierarchy of the new Yukawa couplings $y_\psi$, $y_{\chi\psi}$ and $y_\chi$ is not explained, its origin could be related to the SM flavour puzzle.  Although we focus on this particular region of parameter space to highlight the importance of finite temperature effects, which will prove to dramatically alter the DM relic abundance, we note that these effects may still be numerically important in much wider regions of parameter space.

\begin{table}
  \centering
  \begin{minipage}{\textwidth}
    \begin{ruledtabular}
    \begin{tabular}{cccccccc}
           P.T. & Mechanism     & $\lambda_p$  & $y_\psi$ & $m_S(T=0)$ & $y_{\chi\psi}$ &  $\chi$ decay ended by & Section \\
      \hline
      \vspace{-0.4cm}\\
      One--step & 
      Instantaneous f.o. & 
      $\ll 1$&
       $>0$&
       $0.1 \text{ -- } 10\GeV$&
       $ \sim 10^{-7}$&
       $\vev{S}$ &
      \ref{sec:instantaneous} 
       \\
      One--step & 
      Decaying DM& 
      $\ll 1$&
      $>0$&
       $0.1 \text{ -- } 10\GeV$&
      $ \sim 10^{-8}$&
       $\vev{S}$&
      \ref{sec:one-step-decay} 
       \\
      \vspace{-0.3cm}\\
      Two--step &
      Instantaneous f.o.&
      $\sim1$ &
      $<0$&
       $\sim100\GeV$&
      $  \sim10^{-6}$&
      vev flip-flop& 
      \ref{sec:two-step-decay} 
      \\
      Two--step & 
      Decaying DM&
      $\sim1$& 
      $<0$&
       $\sim100\GeV$&
      $ \sim 10^{-7}$&
      vev flip-flop & 
      \ref{sec:two-step-decay} 
      \\
      \bottomrule
    \end{tabular} 
    \end{ruledtabular}
  \end{minipage}
  \caption{A summary of the mechanisms that operate in various regions of parameter space.  
  In the first column we give the nature of the phase transition of the new scalar $S$.  
  In the penultimate column we note whether the $\chi$ decay and inverse--decay channel
  closes due 
  to the vev of $S$ becoming sufficiently large, or to the vev of $S$ becoming 
  zero after a `vev flip-flop'.
  }
  \label{tab:summary}
\end{table}

\section{The Effective Potential at Finite Temperatures}
\label{sec:effpot}

We first turn to the effective potential of the scalar fields, and 
describe the finite temperature corrections we include.
The effective potential is analogous to the free-energy of a system, 
and the system (in this case the vacuum scalar field configuration) 
will move to the minimum of the effective potential. 
The leading terms in the zero-temperature effective potential are
the tree-level potential $V^\text{tree}(h,S)$, 
the one-loop Coleman-Weinberg correction $V^\text{CW}(h,S)$~\cite{Coleman:1973jx}, 
and a one-loop counterterm $V^\text{CT}(h,S)$.
The leading corrections at finite temperature are
the one-loop thermal corrections $V^T(h,S,T)$~\cite{Dolan:1973qd}
and the resummed higher order ``daisy'' diagrams~\cite{Carrington:1991hz,
Quiros:1999jp, Delaunay:2007wb}.  The one-loop effective potential 
at finite temperature, $V^\text{eff}(h,S,T)$, is then
\begin{align}
  V^\text{eff}(h,S,T) &= V^\text{tree}(h,S)  +  V^\text{CW}(h,S)
                       + V^\text{CT}(h,S)
                       + V^T(h,S,T) + V^\text{daisy}(h,S,T)\,.
  \label{eq:app:Veff}
\end{align}
The tree level potential $V^\text{tree}$ is given in \cref{eq:kin:V}.  
The $T$-independent Coleman-Weinberg contribution
is~\cite{Coleman:1973jx, Quiros:1999jp}
\begin{align}
  V^\text{CW}(h,S) &=
    \sum_i \frac{n_i}{64 \pi^2} m_i^4(h,S)
    \bigg[ \log\bigg( \frac{m_i^2(h,S)}{\Lambda^2} \bigg)
               - C_i \bigg] + V^\text{CT}(h,S) \,,
  \label{eq:app:VCW}
\end{align}
where the sum is over the eigenvalues of the field-dependent mass matrices 
of all fields which couple significantly to the scalars and 
$|n_i|$ accounts for their respective numbers of
degrees of freedom. 
In our case $n_h = n_{G^0} = n_S = 1$, $n_{G^+} = 2$, $n_Z = 3$, $n_W = 6$ and $n_t = -12$.  
We do not include the lighter quarks as they couple only very weakly to the SM Higgs.  
The coefficient $n_i$ is positive for bosons and negative for fermions, due to their different statistics.  
As a renormalisation scale $\Lambda$ we use the 
characteristic scale of the $S$ field, $\mu_S$.  
Using the dimensional regularisation scheme, $C_i = 5/6$ for gauge
bosons and $C_i = 3/2$ for scalars and fermions.  
The field-dependent masses of  
the CP even neutral scalars, in the 
basis $(h, s)$,  are
\begin{align}
m_{hS}^2 =&\,
  \begin{pmatrix}
    -\mu_H^2 + 3 \, \lambda_H \, h^2 + \frac{1}{2}\lambda_p \, S^2  &  \lambda_p \, h \, S \\
    \lambda_p \, h \, S   &   -\mu_S^2 + \frac{1}{2} \lambda_{S} \, S^2 +\frac{1}{2}\lambda_p \, h^2 
  \end{pmatrix},
  \label{eq:m2-neutral}
\end{align}
while the field dependent masses of the remaining bosons are
\begin{align}
  m_{G}^2   &= -\mu_H^2 + \lambda_H \, h^2 + \frac{\lambda_p \, S^2}{2} \,,\\
  m_{W^\pm}^2 &= \frac{1}{4} g^2 \, h^2  \,,\\
  m_{Z}^2     &= \frac{1}{4}(g^2 + g'^2) \, h^2 \,,\\
  m_\gamma^2  &= 0 \,, \label{eq:m2-gamma} \\
  m_t^2       &= \frac{1}{2} y_t^2 \, h^2 \,,
\end{align}
where $g'$ and $g$ are the SM $u(1)_Y$ and $su(2)_L$ coupling 
constants, respectively. 
Although the dark sector fermion $\psi$ couples strongly to $S$, 
we neglect its contribution since 
its mass is always much larger 
than either the temperatures or the scalar field values we consider~\cite{Comelli:1996vm}.

To ensure that $\langle H \rangle = \mu_H/\sqrt{\lambda_H}$, $m_h^2 = 2 \mu_H^2$ and that $m_S$ 
is given by its tree level value at $T=0$, we add the counterterm
\begin{align}
  V^\text{CT}(h,S) &=
    - \frac{1}{2}\delta_\mu h^2
    + \frac{1}{4}\delta_\lambda h^4 -
      \frac{1}{2}\delta_{\mu_S} S^2,
  \label{eq:app:VCT}
\end{align}
where the factors $\delta_i$ are
\begin{align}
  \delta_\mu &=
    \frac{3}{2 v} \be{\frac{\partial V^\text{CW}}{\partial h}}{v} 
  - \frac{1}{2}\be{\frac{\partial^2 V^\text{CW}}{\partial h^2}}{v}\,, \\
  \delta_\lambda &=
    \frac{1}{2 v^3} \bigg(\be{\frac{\partial V^\text{CW}}{\partial h}}{v}
  - v \be{\frac{\partial^2 V^\text{CW}}{\partial h^2}}{v} \bigg) \,,\\
  \delta_{\mu_S} &=
    \be{\frac{\partial^2 V^\text{CW}}{\partial S^2}}{v}\,.
\end{align}
The one-loop finite temperature correction is~\cite{Dolan:1973qd}
\begin{align}
  V^T(h,S,T) &= \sum_i \frac{n_i T^4}{2\pi^2}\int_0^\infty \! dx \, x^2
              \log \bigg[ 1 \pm
                \exp\Big( -\sqrt{x^2+m_i^2(h,S)/T^2} \Big) \bigg] \,,
  \label{eq:app:VT}
\end{align}
where again we sum over the same eigenvalues as in \cref{eq:app:VCW}.  
The positive sign in the integrand is for fermions and the negative sign is for bosons.

Finally, higher order diagrams containing the bosons give rise to 
the so-called ``daisy'' corrections~\cite{Carrington:1991hz, Quiros:1999jp, Ahriche:2007jp,
Delaunay:2007wb}
\begin{align}
  V^\text{daisy}(h,S,T) &= -\frac{T}{12\pi} \sum_i n_i
    \Big( \left[ m^2(h,S) + \Pi(T) \right]_i^\frac{3}{2}
    - \left[ m^2(h,S) \right]_i^\frac{3}{2}
  \Big) \,,
  \label{eq:app:Vdaisy}
\end{align}
where the first term should be interpreted as the $i$-th eigenvalue of the
matrix-valued quantity $[m^2(h,S) + \Pi(T)]^{3/2}$. Here, $m^2(h,S)$ is the
block-diagonal matrix composed of the individual mass matrices~\cite{Patel:2011th}.
This sum runs only over the bosonic degrees of freedom.
The thermal (Debye) masses~\cite{Carrington:1991hz} in \cref{eq:app:Vdaisy} are
\begin{align}
  \Pi_{h,G^0, G^+}  &= \tfrac{1}{48} T^2 \left(9 g^2 + 3g'^2
                     + 24\lambda_H + 12 y_t^2 + 2\lambda_p\right) \,,\\
  \Pi_{S}  &= \tfrac{1}{24} T^2 \left( 4 \lambda_p +  \lambda_S \right) \,,\label{eq:PiS}\\
  \Pi_{W^{1,2,3}}^L &= \tfrac{11}{6} g^2 T^2 \, ,\\
  \Pi_{W^{1,2,3}}^T &= 0 \,,\\
  \Pi_B^L           &= \tfrac{11}{6} g'^2 T^2 \,, \\
  \Pi_B^T           &= 0 \,,
\end{align}
where $\Pi_{h, G^0, G^+}$ are the thermal masses of the components of $H$,
$\Pi_S$ is the thermal mass of the components of $S$ 
and $\Pi_{W,B}$ are the thermal masses of the electroweak gauge boson.  
Only the longitudinal components of the gauge
bosons obtain a non-zero thermal mass.

\subsection{One-step Phase Transition}
\label{sec:one-step}


\begin{figure}
  \begin{center}
    \begin{tabular}{cc}
      \includegraphics[width=0.45\textwidth]{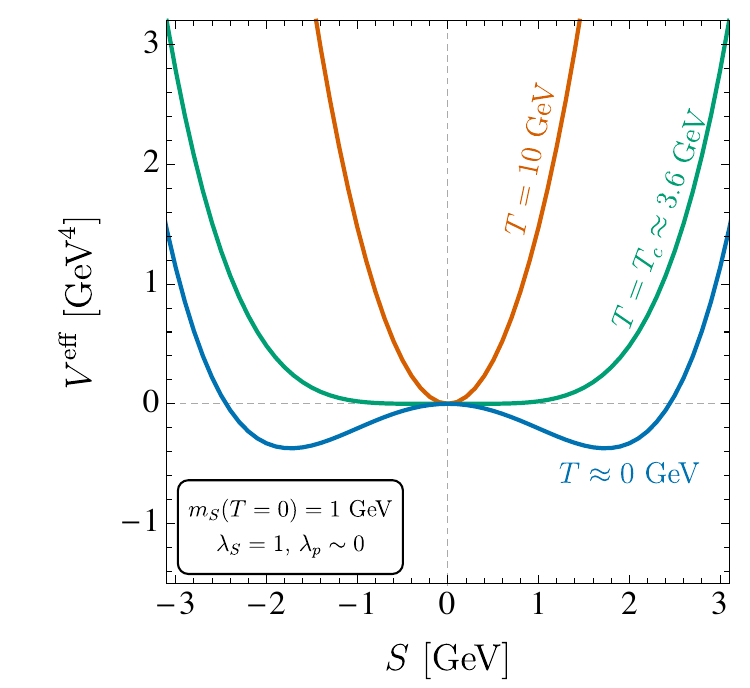} & 
      \includegraphics[width=0.45\textwidth]{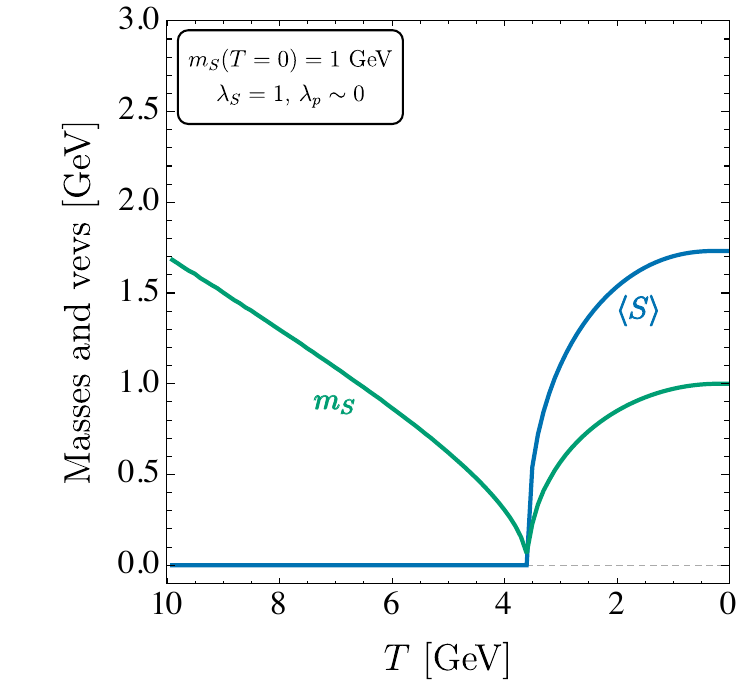}
    \end{tabular}
  \end{center}
  \caption{Left:~the effective potential as a function of the $S$ field value at a high temperature ($10\GeV$), at the critical temperature $T=T_c\approx3.6\GeV$ 
  and at $T\approx0\,\text{GeV}$.  We subtract a constant term at each $T$ so that
   $V^\text{eff}(0)=0$.  
   Right:~the scalar mass and vev as a function of temperature.
   }
  \label{fig:one-step-eff-pot}
\end{figure}


With the effective potential in hand, we can now consider its behaviour in 
various regions of parameter space.  We will first consider the regime where 
$\lambda_p \ll 1$, where the new scalar $S$ and the SM Higgs are  
weakly coupled.  In this case the effective potential in $h$ and $S$ 
decouples, $V^\text{eff}(h,S,T) = V^\text{eff}(h,T) + V^\text{eff}(S,T)$.
To consider the evolution of the new scalar, it is sufficient to 
consider the effective potential as a function of $S$ alone.

In \cref{fig:one-step-eff-pot} (left) we show $V^\text{eff}(S,T)$ 
at several temperatures, for a particular choice of parameters.  
We will be interested in $\mu_S > 0$, so that 
$S$ obtains a vev at $T=0$.  
We see that at high temperatures, thermal corrections dominate the 
effective potential and the minimum is at $S=0$, so there is no vev.  As the universe expands, the temperature reduces until 
$T\simeq \mu_S$, at which point the finite $T$ 
corrections become similar in size to the $T=0$ potential.  
At $T_c$ there is a second-order phase transition and $S$ obtains a vev.  
As the universe cools further, the minima deepen and the vev increases 
to around $1.7\GeV$.

As well as the position of the vev, the effective potential also determines the 
physical mass of the scalar $S$ in the thermal bath.  
Finite temperature corrections at one-loop are taken into account 
by taking the second derivative of the effective potential at a minimum.
In \cref{fig:one-step-eff-pot} (right) 
we show the evolution of the vev along with the physical $S$ mass, as a function 
of temperature.  We see that the mass is large at high temperatures, 
becomes small through the phase transition (at $T_c$ the second derivative 
of the effective potential goes to zero), and reaches a value $\simeq \mu_S$ 
at zero temperature.

In \cref{sec:instantaneous,sec:one-step-decay} we will show that this 
temperature dependence of the mass and vev can lead to kinematic thresholds 
opening and closing, 
allowing dark matter to come into equilibrium or decay in some temperature window.

\subsection{Two-step Phase Transition}
\label{sec:two-step}


\begin{figure}
  \begin{center}
    \begin{tabular}{c}
      \includegraphics[width=0.7\textwidth]{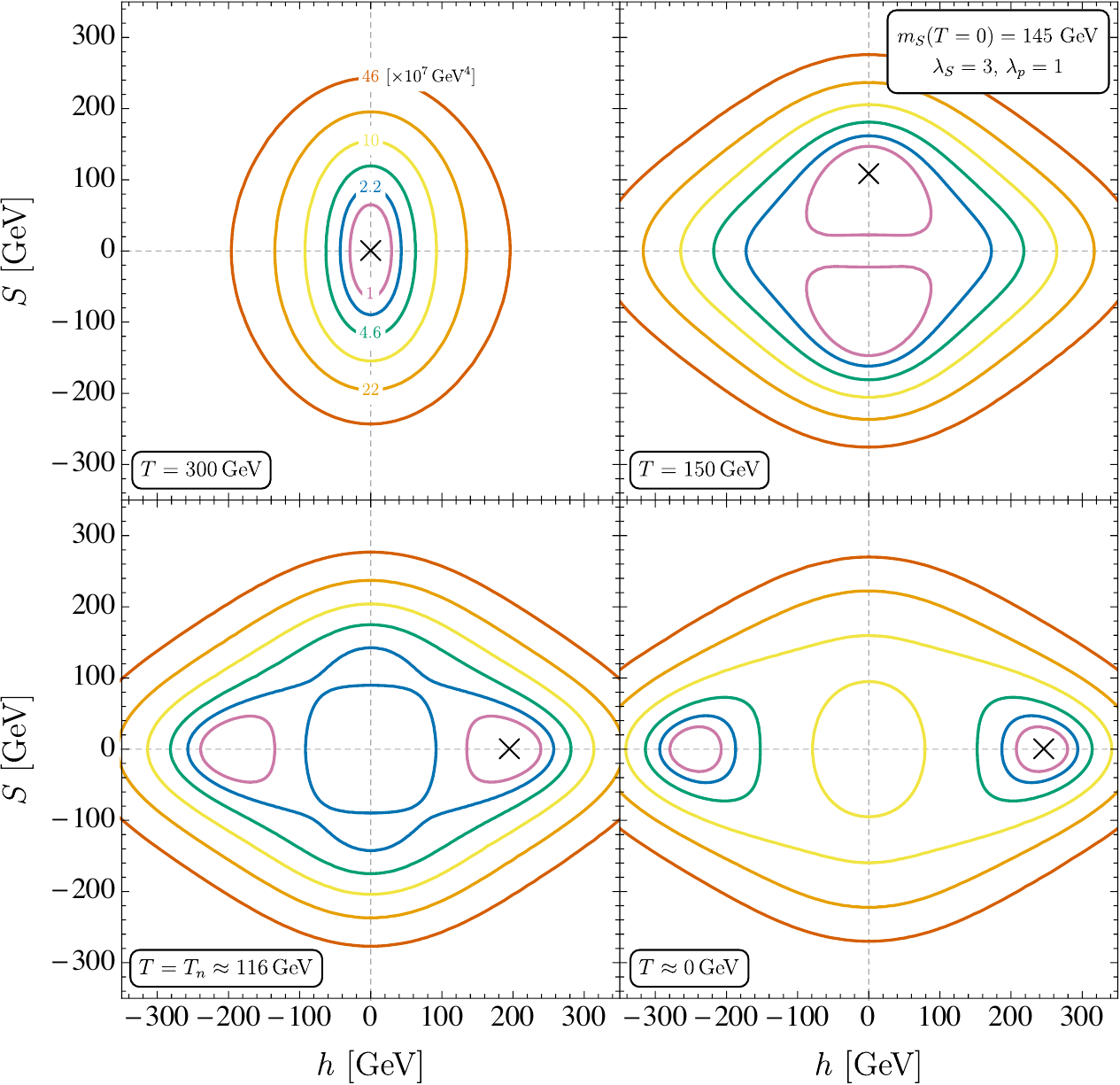}
    \end{tabular}
  \end{center}  
  \caption{The effective potential $V^\text{eff}$ at $T=300\GeV$ (top-left), 
  at $T=150\GeV$ (top-right), the nucleation temperature $T=T_n \approx 116\GeV$ (bottom-left) 
  and $T\approx0\GeV$ (bottom-right).  
  In each plot we subtract a constant, the minimum value of the potential,  
  to highlight the features around the minima.  
  The black cross indicates the phase the Universe is
  in at the given temperatures.  
  }
  \label{fig:two-step-eff-pot}
\end{figure}



\begin{figure}
  \begin{center}
    \begin{tabular}{c}
      \includegraphics[width=0.45\textwidth]{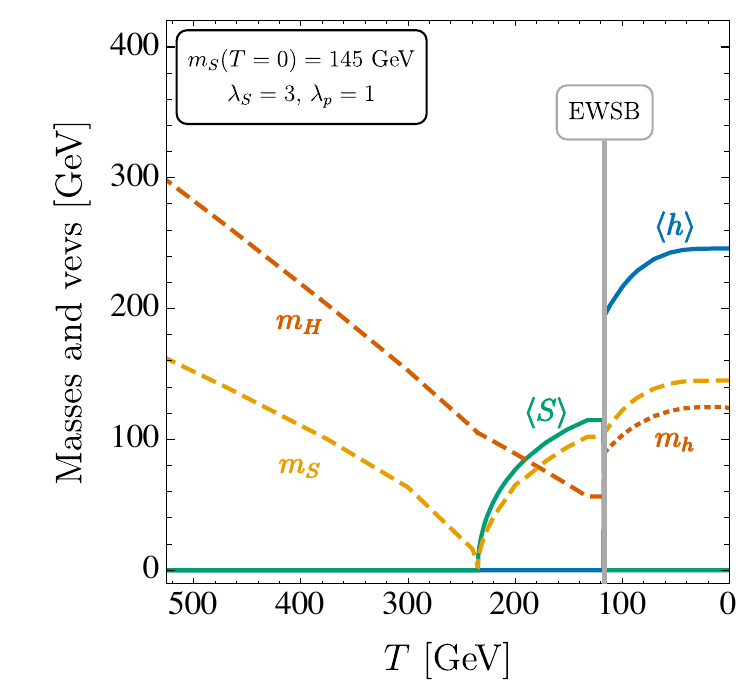}
    \end{tabular}
  \end{center}  
  \caption{
  Evolution of the scalar vevs and masses with temperature
    for a particular parameter point.
    $m_H$ denotes the mass of the SM Higgs doublet above electroweak symmetry breaking,
    while $m_h$ is the mass of the SM-like physical Higgs boson below.
    The interaction between the SM Higgs and the new scalar, $S$, has 
    reduced the EWSB temperature from its usual value around $160\GeV$ to $116\GeV$.
  }
  \label{fig:two-step-masses-vevs}
\end{figure}


In the previous section we assumed the new scalar and 
the SM Higgs to be weakly coupled. 
If instead they are 
coupled with $\lambda_p \approx 1$, we must consider 
the effective potential as a function of both $S$ and $h$.  
We are in particular interested in the region of parameter space 
which exhibits a two-step phase transition (also called a vev flip-flop).  
In \cref{fig:two-step-eff-pot} we show an example of 
such a transition.
If $\mu_S \approx \mu_H$, it can happen that at high temperatures 
(top-left) there is one minimum at $(h,S) = (0,0)$ and neither field has 
a vev.  As the temperature reduces, the finite temperature corrections 
reduce and minima develop at $\vev{S}\neq0$ due to the $T=0$ potential.  
As the temperature drops further, further minima appear at $\vev{h}\neq0$, 
(top-right).  
At this point there is a barrier between the $\vev{S}\neq0$ minima and the 
$\vev{h}\neq0$ minima, and there is a period of supercooling while 
the universe remains in this meta-stable phase.
As the temperature drops further, a first order phase transition may 
take place, when the formation and growth of bubbles of the new 
phase is energetically favourable~\cite{Linde:1981zj}.
We calculate the temperature of the phase transition (here named the nucleation temperature) 
using the publicly 
available code \texttt{cosmoTransitions}~\cite{Wainwright:2011kj,Kozaczuk:2014kva,Blinov:2015sna, Kozaczuk:2015owa}. 
Note that away from the minima the effective potential is gauge dependent, 
so the nucleation temperature in principle has a residual gauge dependence~\cite{Patel:2011th}, 
which we neglect.
At the nucleation temperature $T_n$, the universe passes to the phase 
where $\vev{h}\neq0$ and $\vev{S}=0$ (bottom-left).  As the temperature reduces 
further, these minima deepen and the universe ends in a phase with 
$\vev{h}=246\GeV$ and $\vev{S}=0$ (bottom-right).

In \cref{fig:two-step-masses-vevs} we show the scalar masses 
and their vevs as a function of temperature.  
Since at no point do both $h$ and $S$ obtain a vev, there is no 
mixing between them.
As for the one-step phase transition, \cref{sec:one-step}, the scalars have masses similar 
to $T$ at high temperature.  In this example, $S$ obtains its vev 
in a second order phase transition at $T \approx 230 \GeV$.  
As in the one-step phase transition, the mass becomes very small during this 
transition.  After this first transition, $S$ obtains a vev, which grows as the 
temperature reduces.  The mass of $S$ starts to grow while the Higgs   
continues to get lighter.  This continues until the nucleation temperature 
$T_n = 116\GeV$, when the universe passes to the phase where 
$\vev{S}=0$ and $\vev{h}\neq 0$ in a strongly first-order phase transition.  
After this second phase transition, which breaks electroweak symmetry, 
three components of the Higgs doublet are eaten by the gauge bosons, 
and the mass of the remaining scalar grows until it reaches $125\GeV$ 
at $T=0$.  Similarly, the $h$ vev grows to $246\GeV$ at $T=0$.

\subsection{Phase Diagram}


\begin{figure}
  \begin{center}
    \begin{tabular}{cc}
      \includegraphics[height=0.45\textwidth]{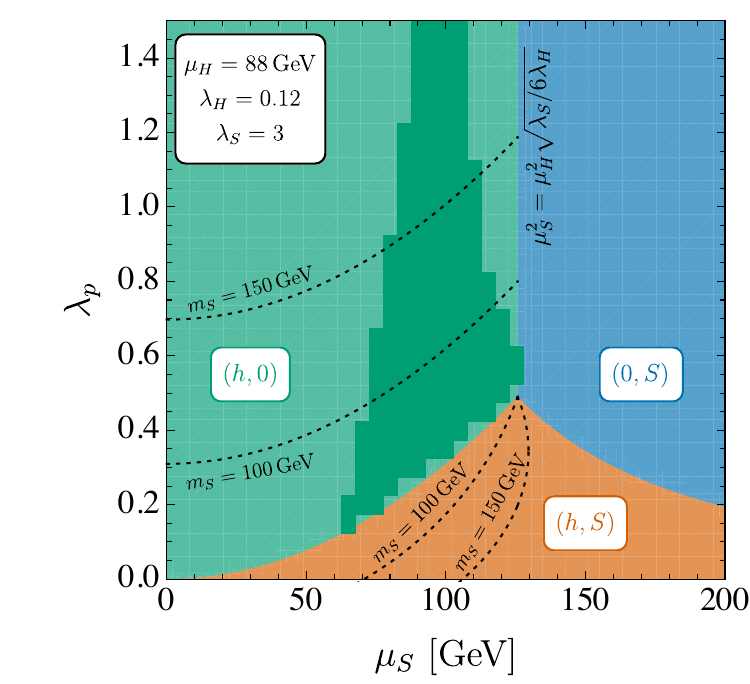} & 
      \includegraphics[height=0.45\textwidth]{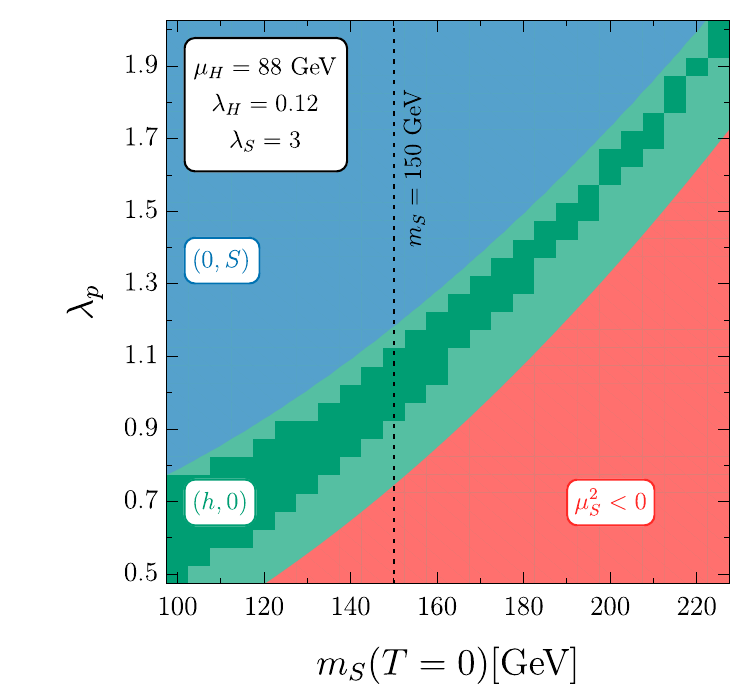}
    \end{tabular}
  \end{center}
  \caption{Left:~the phase diagram showing which fields obtain a vev in the 
  global minimum of the tree-level effective potential 
  at $T=0$, as a function of $\mu_S$ and $\lambda_p$.  Also shown is $m_S(T=0)$ 
  for the physical $(h,0)$ and $(h,S)$ minima.  In $(0,S)$ electroweak symmetry is not broken, 
  so it is not physical. 
  Right:~the region where a two-step phase transition occurs (green) along with 
  regions where $S$ never obtains a vev (red) and where electroweak 
  symmetry is not broken at $T=0$ (blue) as a function of $m_S(T=0)$ and $\lambda_p$.
  Superimposed on both in dark green is the region where \texttt{cosmoTransitions} gives a two-step phase transition 
  for the one-loop effective potential.
   }
  \label{fig:phase-diagram}
\end{figure}


Now that we have outlined the two main scenarios of interest, we 
briefly discuss the parameter space of the effective potential.  
In \cref{fig:phase-diagram} (left) 
we show the phase diagram of the tree-level potential for $\lambda_S = 3$.  
The parameter space is  
divided into regions where at zero temperature the global minimum 
is one with 
(green) $(\vev{h} \neq 0, \vev{S} = 0)$, 
(orange) $(\vev{h} \neq 0, \vev{S} \neq 0)$ and
(blue) $(\vev{h} = 0, \vev{S} \neq 0)$. 
The blue region does not correspond to our universe, 
since electroweak symmetry is broken, 
but either the green or orange region may be physical.  
In these regions we also plot contours of the $S$ mass at $T=0$.  
The orange region shows where a one-step phase transition occurs, discussed in 
\cref{sec:one-step}.  
To simplify calculations we later
assume $\lambda_p \ll 1$, but the mechanisms we discuss would be seen in 
the whole orange region.

A two-step phase transition, discussed in 
\cref{sec:two-step}, may occur in the green region.
The pixellated region shows points where   
\texttt{cosmoTransitions} finds that $S$ obtains a vev at some temperature, 
and then at a later lower temperature, 
as the SM Higgs obtains its vev, the $S$ vev goes back to zero 
(also called a vev flip-flop) for the one-loop effective potential.
Here, 
the vev of $h$ gives a contribution to the mass of $S$ and the relation 
between $m_S(T=0)$ and $\mu_S$ is 
\begin{align}
m_S^2(T=0) =&\, \frac{\lambda_p}{2} \, \langle H \rangle^2 - \mu_S^2 \,.
\end{align}
We see that when, e.g., $m_S(T=0) = 150\GeV$, only certain 
values of $\lambda_p$ are allowed.  This is seen to correspond to 
\cref{fig:phase-diagram} (right), where we again plot the green region, 
now on the $m_S(T=0)$ -- $\lambda_p$ plane. 
In the lower-right red region, which shows where $\mu_S^2 < 0$ at tree-level, 
the $S$ symmetry is 
never broken and $S$ never obtains a vev. 
In the blue region the deepest minima occur when $S$ has a vev 
and $h$ has no vev, so electroweak symmetry is not broken 
and this region does not correspond to 
our universe.  
The pixellated region again shows where \texttt{cosmoTransitions} calculates that 
the two-step phase transition successfully completes 
for the one-loop effective potential.  
As $\lambda_S$ increases, 
the two-step phase transition region becomes larger.  For a given $m_S(T=0)$, 
the $S$ vev and the temperature at which $S$ first obtains a vev 
both increase with $\lambda_S$.
Although not shown in these diagrams, $S$ obtains a vev at higher $T$ for larger portal couplings.

\section{Finite Temperature Corrections and the Thermal Bath}
\label{sec:thermal-bath}

We now turn to a discussion of the thermal bath.  
In the hot, early Universe, there is enough energy to produce 
 particles with masses $\lesssim T$.  If there are efficient processes 
 which create and destroy a particle, it will come into equilibrium with 
 a number density $n \sim T^3$.  In this work, we will assume that 
 after inflation and above $T \simeq 100\TeV$, there were efficient 
 processes which led to $\chi$, $\psi$ and $S$ to come into 
 equilibrium with the thermal bath.  For freeze-in 
 scenarios where $\chi$ does not thermalise in the early universe, 
 see~\cite{Baker:2017zwx}.

To determine if certain processes are efficient at keeping 
the new particles in contact with the thermal bath as the temperature cools, 
we must calculate the rates of these processes and compare them 
to the rate of Hubble expansion.  
The masses which 
enter the Feynman rules are given by the imaginary part 
of the self energies, which are modified in the presence of 
thermal corrections~\cite{Weldon:1983jn,Carrington:2002bv}.  
Although both bosons and fermions obtain these thermal corrections, 
their different statistics lead to different boundary conditions 
in the compactified dimension at finite temperature.  The 
bosonic contributions contain a Matsubara zero-mode while the fermionic 
contributions do not, so
 the fermionic contributions are subleading to the bosonic 
contributions.  We therefore only include 
the thermal corrections to the boson masses. 
 However, from \cref{eq:kin:L}, we can see that when $S$ has a 
 vev, the effective mass parameter for $\psi$ will receive an extra contribution,
 \begin{align}
 m_\psi(T) =&\, \tilde{m}_\psi + y_\psi \vev{S}(T)\,,
 \end{align}
so $m_\psi$ will still depend on temperature.
Since we imagine $y_\chi$ to be small, we will take 
$m_\chi$ to be temperature independent.  
Since we will mostly be interested in small mass splittings, we 
introduce the parameter
\begin{align}
\Delta =&\, \frac{m_\chi - \tilde{m}_\psi}{m_\chi}\,.
\end{align}

These temperature 
dependent masses mean that kinematic thresholds can open 
or close as the temperature reduces.
In this work, we focus on scenarios where decay and inverse decay of $\chi$, 
which is only allowed when 
$m_\chi > m_\psi(T) + m_S(T)$, 
has a dramatic impact on the resulting dark matter relic abundance.  
In \cref{sec:instantaneous} we consider 
a scenario where $\chi$ remains in equilibrium until the threshold 
closes, at which point there are suddenly no processes which keep 
$\chi$ in contact with the thermal bath and it immediately freezes-out.  
We call this process \emph{instantaneous freeze-out}.
In this case, the relic abundance is set by the temperature at which 
the threshold closes, and $m_\chi \simeq 30 \,T_\text{End}$ results 
in the observed relic abundance.
In \cref{sec:one-step-decay} we consider a scenario where 
the kinematic threshold is closed at high temperatures and 
so $\chi$ freezes-out when relativistic. Then, as the finite temperature 
corrections to $m_S$ become smaller,
the threshold opens and $\chi$ can decay for some time.  
Finally, 
the mass of $\psi$ increases due to $S$ obtaining a vev, closing the threshold 
and stabilising $\chi$, which we call \emph{decaying dark matter}.
The final relic abundance will then be determined 
by the amount of decay that has occurred when the threshold closes.
The abundance of $\chi$ will approach but 
not reach equilibrium and produce the observed relic abundance.  
Ensuring that $\psi$ and $S$ remain in equilibrium 
means that $\chi\to\psi S$ efficiently depletes the abundance of $\chi$, 
with the energy density being passed to the thermal bath.
In \cref{sec:two-step-decay} we consider a similar scenario, but where $S$ is strongly 
coupled the SM Higgs and a two-step phase transition can occur.  
In this case, the mass of $\psi$ reduces when $S$ obtains a vev 
and the kinematic threshold opens, 
allowing $\chi$ to decay, 
and closes when the $S$ vev disappears.  
Again the abundance of $\chi$ will approach but 
not reach equilibrium and can produce the observed relic abundance.  
We call this the \emph{vev flip-flop}.


\begin{figure}
  \centering
  \includegraphics[width=10cm]{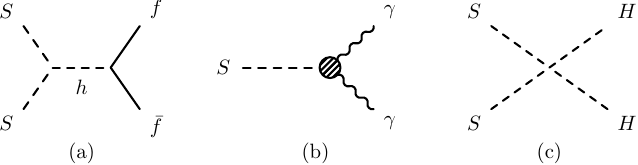}
  \caption{Processes which connect $S$ to the thermal bath.}
  \label{fig:S-diagrams}
\end{figure}


We now consider the constituents of the thermal bath.  
The new particles we introduce do not change 
the evolution of the SM particles, which follow 
the standard cosmology.  
For the new particles, we will ensure that 
$S$ and $\psi$ remain in equilibrium when number changing processes of $\chi$ are active.  Although 
this is not always necessary it will simplify our 
calculations.  We will discuss the precise processes 
which keep $S$ and $\psi$ in equilibrium in each section, 
and here give an overview.

In \cref{fig:S-diagrams} we show three possible ways that $S$ interacts 
with the SM bath.  The processes in \cref{fig:S-diagrams} (a) and (b) can only keep 
$S$ in thermal equilibrium when electroweak symmetry (EWS) is broken,
while process (c) is always active.   
Process (a) and (b) will mostly be relevant in 
 \cref{sec:instantaneous,sec:one-step-decay}.
Process (c) 
will keep $S$ in thermal equilibrium in \cref{sec:two-step-decay}, 
where EWS is not broken during $\chi$ decay.
The annihilation cross sections and decay rates for the processes in
\cref{fig:S-diagrams} are
\begin{align}
  \sigma(S S \to f \bar{f})
    &= \frac{C_f \, \lambda_p^2 \, y_f^2 \, v^2 }{64\pi s \sqrt{s - 4 m_S^2}}
       \frac{ (s - 4 m_f^2)^{3/2}}
                        {(s - m_h^2)^2
                        }\,,
  \label{eq:S-S-f-f} \\[0.2cm]
  \Gamma(S  \to \gamma \gamma)
    &= \frac{v^2 }{16 \pi m_S}
    \left(\frac{ \lambda_p \,v \vev{S}}{m_h^2 - m_S^2}\right)^2
       |F_W^{\gamma\gamma}+F_f^{\gamma\gamma}|^2 \,,
  \label{eq:kin:S-gamma-gamma}\\[0.2cm]
  \sigma(S S \to H H)
    &= \frac{\lambda_p^2 }{256 \pi s}
       \sqrt{\frac{ s - 4 m_H^2}
                        {s - 4 m_S^2}} \,,
  \label{eq:kin:S-S-H-H} 
\end{align}
where $C_f$ is a colour factor, $y_f = m_f \,\sqrt{2} / v$ 
is the Yukawa coupling of the SM fermions 
to the Higgs, $v$ is the SM Higgs vev, 
and the factors $F_W^{\gamma\gamma}$ and $F_f^{\gamma\gamma}$ 
are given in~\cite{Kopp:2013mi,Djouadi:2005gi}.


\begin{figure}
  \centering
  \includegraphics[width=10cm]{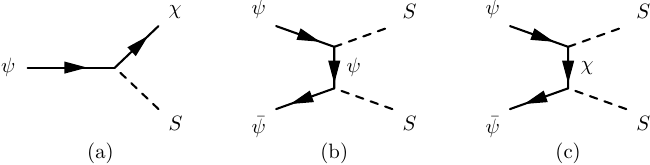}
  \caption{Processes which connect $\psi$ to the thermal bath.}
  \label{fig:psi-diagrams}
\end{figure}



In \cref{fig:psi-diagrams} we show processes that may keep 
$\psi$ in contact with the thermal bath.  Since we will be interested 
in situations where $\chi$ is out of equilibrium, we will choose values of $y_\psi$ so that 
$\psi$ can remain in equilibrium through \cref{fig:psi-diagrams} (b) 
until the abundance of $\chi$ is fixed.
We also want $\chi$ to be the dominant dark matter relic, 
so we will choose $y_\psi$ large enough that 
$\psi$ freezes-out with less than 10\% to the relic abundance.
To ensure this, we will need 
$y_\psi \gtrsim 1$ in \cref{sec:instantaneous,sec:one-step-decay} and 
$|y_\psi| \gtrsim 4$ in \cref{sec:two-step-decay}.  
This large $y_\psi$ also has the benefit of providing 
a large mass contribution to $\psi$, when $S$ obtains a vev.
Process (c) goes as $y_{\chi\psi}^4$ and will be too small to keep $\psi$ in equilibrium.
In what follows, we will see that $m_\psi(T)$ is always smaller than $m_\chi + m_S(T)$, so 
the decay process in (a) will always be kinematically forbidden and $\psi$ will only 
decay via an off-shell $S$.  
The annihilation cross section for the processes in
\cref{fig:psi-diagrams} (b) is
\begin{align}
  \sigma(\psi \bar{\psi} \to S S)
    &\simeq \frac{y_\psi^4 }{64\pi s^2 (s - 4 m_\psi^2)}
      \bigg[ 2 (s^2 + 16 m_\psi^2 s - 32 m_\psi^4)
        \log\bigg( \frac{s + \sqrt{s (s - 4 m_\psi^2)}}
                        {s - \sqrt{s (s - 4 m_\psi^2)}} \bigg)
                                \nonumber\\
    &\hspace{5cm}
      - 4 (s + 8 m_\psi^2) \sqrt{s (s - 4 m_\psi^2)} \bigg] \,,
  \label{eq:kin:sigma-psi-psi-S-S}
\end{align}
where we have taken the limit $m_S \to 0$.  
We keep the full $m_S$ dependence in our numerical work.


\begin{figure}
  \centering
  \includegraphics[width=10cm]{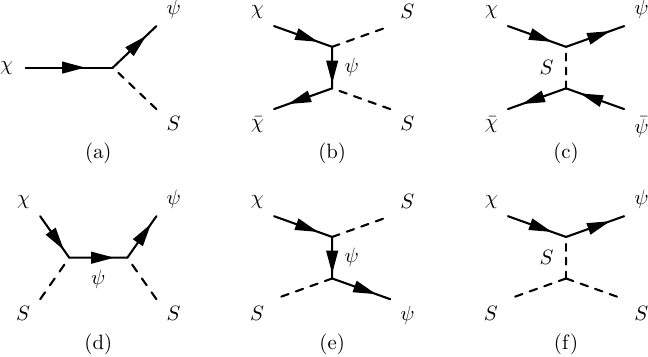}
  \caption{Processes which may connect $\chi$ to the thermal bath.  
  We do not show diagrams involving $\lambda_p$, please see text.}
  \label{fig:chi-diagrams}
\end{figure}



Finally, $\chi$ may be in equilibrium through the processes 
shown in \cref{fig:chi-diagrams}.
The decay and inverse decay diagram, \cref{fig:chi-diagrams} (a), 
is proportional to $y_{\chi\psi}^2$, while the processes 
\cref{fig:chi-diagrams} (b) and (c) are at least proportional to $y_{\chi\psi}^4$.
We will choose $y_{\chi\psi}$ small enough that the processes proportional to 
$y_{\chi\psi}^4$ do not come into equilibrium at the temperatures of interest.
The decay rate of the process in \cref{fig:chi-diagrams} (a) is
\begin{align}
  \Gamma(\chi \to \psi S)
    &= \frac{y_{\chi\psi}^2}{16 \pi}
       \frac{(m_\chi + m_\psi)^2 - m_S^2}{m_\chi^3}
       \sqrt{ \big[ m_\chi^2 - (m_\psi + m_S)^2 \big]
              \big[ m_\chi^2 - (m_\psi - m_S)^2 \big] } \,.
  \label{eq:kin:chi-psi-S}
\end{align}
Note that we do not show $\chi \psi \leftrightarrow S$.  Although this channel 
is kinematically open in the very early universe, when 
$m_S(T) > m_\chi + m_\psi(T)$, we always take 
$m_\chi, \tilde{m}_\psi \approx (30 -100) \, m_S(T=0)$, so this channel 
will be closed at the temperatures where the abundance of $\chi$ is being set. 
We also do not show $\psi \leftrightarrow \chi S$ since for the parameter points we consider 
this decay is always kinematically forbidden.  

In (d) - (f) we show channels contributing to $\chi S \leftrightarrow \psi S$.  
The cross section for the process is
\begin{align}
  \sigma(S \chi \to S \psi)
    &\simeq \frac{y_\psi^2 y_{\chi\psi}^2}{32\pi s^2 (s - m_\chi^2)^3}
       \Big[ (s - m_\chi^2) (5 s^3 + 55 m_\chi^2 s^2 + 3 m_\chi^4 s + m_\chi^6)
                                \nonumber\\
    &\hspace{5cm}
           - 2 s^2 ( s^2 - 18 m_\chi^2 s - 15 m_\chi^4) \log\Big( \frac{m_\chi^2}{s} \Big)
       \Big] \,,
  \label{eq:kin:sigma-S-chi-S-psi}
\end{align}
where we have taken the limit $m_\psi \simeq m_\chi$
and $m_S \simeq 0$. We have also set $v_S = 0$ since (f) is 
subdominant to (d) and (e) for the parameters we consider. 
In our numerical
analysis we use the full expressions.  We do however always take 
$\Gamma_\psi = \Gamma_S = 0$ since the widths of $\psi$ and $S$ are very small.
At temperatures near $m_S$ but lower than $m_\chi$,  
$n_S \langle \sigma(S \chi \to S \psi) v \rangle \sim T^3/ m_\chi^2 \sim m_\chi / x^3$ where $x = m_\chi / T$, hence the rate of this process will generally be suppressed by a factor 
$x^3$ compared to $\Gamma(\chi \to \psi S) \sim m_\chi$. 
However, the rate can be resonantly enhanced when $m_S(T) \approx 0$.

Processes such as 
 $\chi \psi \to S \to S S S$ and $\chi \psi \to S \to S S$ (when $S$ has a vev) 
 are also possible, although the rates will be small as the abundance 
 of $\psi$ will be Boltzmann suppressed at the temperatures of interest, 
 and the intermediate $S$ will be a long way off-shell.
We note that although there is in principle 
mixing between $\chi$ and $\psi$ when $\vev{S} \neq 0$,
this mixing will be small and we will ignore its negligible effects. 
We also neglect diagrams involving $\lambda_p$. This coupling is taken to be small in 
\cref{sec:instantaneous} and \cref{sec:one-step-decay}, while in 
\cref{sec:two-step-decay} these processes lead to a rate significantly smaller than the Hubble rate, 
due to the $x^3$ suppression mentioned above.

\section{Dark Matter Abundance and the Boltzmann Equations}
\label{sec:dmdecay}

We will be interested in the abundance of $\chi$  
whether or not it is in equilibrium so we will keep track 
of its abundance using Boltzmann equations. 
In general the Boltzmann equation for $\chi$ is
\begin{align}
\dot{n}_\chi + 3Hn_\chi &=  C[n_\chi]\,.
\end{align}
In the parameter space we consider the collision term will have non-negligible contributions from the 
decay process $\chi \leftrightarrow \psi S$ and the scattering 
process $\chi S \leftrightarrow \psi S$,
\begin{align}
C[n_\chi] &= C_{\chi \leftrightarrow \psi S }[n_\chi] + C_{\chi S \leftrightarrow \psi S}[n_\chi] \,.
\end{align}

The collision term for the decay process $\chi \leftrightarrow \psi S$ 
is given by
\begin{align}
 C_{\chi \leftrightarrow \psi S }[n_\chi] = &\, - \int d\Pi_\chi d\Pi_\psi d\Pi_S (2\pi)^4\delta^4(p_\chi-p_\psi-p_S) 
 \times \notag\\
&\hspace{2 cm}
\left( \abs{\mathcal{M}_{\chi \to \psi S}}^2 f_\chi(1\pm f_\psi)(1\pm f_S) - 
\abs{\mathcal{M}_{\psi S \to \chi}}^2 f_\psi f_S (1\pm f_\chi)\right)\,,
\end{align}
where $d\Pi_i$ is the Lorentz invariant phase space of particle $i$ and $f_i$ are their 
phase space densities.
We neglect Pauli-Blocking and Bose-Enhancement, 
i.e., $(1\pm f_i) \approx 1$, and 
assume Maxwell-Boltzmann 
distributions for $\psi$ and $S$.
This is a good approximation for $\psi$, as we will be interested in temperatures much less than the mass 
of $\psi$, but since these temperatures are similar to the mass of 
$S$, the full $S$ distribution should be used for more detailed calculations.
As we argued above, $\psi$ and $S$ will remain in thermal equilibrium, 
and we assume with zero chemical potential,  
during the periods of interest.
These assumptions are useful as, when combined with 
the energy-conservation part of the delta-function, it follows that
\begin{align}
f_\psi f_S = e^{-\frac{E_\psi}{T}}e^{-\frac{E_S}{T}} = e^{-\frac{E_\chi}{T}} = f^{eq}_\chi\,.
\end{align}
Assuming that the decay process is $CP$ invariant, 
the collision term then becomes
\begin{align}
 C_{\chi \leftrightarrow \psi S }[n_\chi] = &\,-\int d\Pi_\chi d\Pi_\psi d\Pi_S (2\pi)^4 \delta^4(p_\chi-p_\psi-p_S)
\abs{\mathcal{M}_{\chi\to\psi S}}^2(f_\chi-f^{eq}_\chi)\nonumber\\
=&\,-2\,m_\chi\int d\Pi_\chi \Gamma_{\chi\rightarrow\psi S} (f_\chi - f^{eq}_\chi)\nonumber\\
=&\,-g_\chi  \int \frac{d^3p_\chi}{(2\pi)^3}\frac{m_\chi}{E_\chi}\Gamma_{\chi\rightarrow\psi S} (f_\chi - f^{eq}_\chi)\,,
\end{align}
where $g_\chi = 2$ is the number of degrees of freedom of $\chi$ 
($\bar{\chi}$ will satisfy an analogous Boltzmann equation, and will also 
contribute towards $\Omega_\chi$).
Note that there is a factor $m_\chi / E = 1/\gamma$.  
It appears since particles with higher momenta experience 
increased time-dilation in the rest frame of the plasma, and so
 their lifetime is increased. Since we will be interested in situations 
 where almost all $\chi$ decay, this time-dilation in 
 the tail gives an important contribution. 
 To solve this equation in practice, we discretise the integral 
 into bins in momentum space and
 track the number density in each bin using
 \begin{align}
n_\chi^i =&\, g_\chi \int_p^{p+\delta p} \frac{d^3 p'}{(2\pi)^3} f_\chi\,,
\intertext{which gives}
  C^i_{\chi \leftrightarrow \psi S }[n^i_\chi] = & - \frac{\Gamma_{\chi\rightarrow\psi S}}{\gamma^i}
(n^i_\chi - n^{eq,i}_\chi)\,,
\intertext{where}
 C_{\chi \leftrightarrow \psi S }[n^i_\chi] = & \sum_i C^i_{\chi \leftrightarrow \psi S }[n^i_\chi]
\end{align}
 For the key results we have checked that changing the number 
 of bins and the highest momentum considered does not significantly 
 change the resulting abundance.  
 
 We follow an analogous derivation for the collision term of the scattering 
process $\chi S \leftrightarrow \psi S$, yielding
 \begin{align}
  C^i_{\chi S\leftrightarrow \psi S }[n^i_\chi] = & - n_S^{eq} \langle \sigma v \rangle 
(n^i_\chi - n^{eq,i}_\chi)\,.
\end{align}
where the thermally averaged cross section for non-identical initial state particles 
is given by
 \begin{align}
 \langle \sigma v \rangle= &\, \frac{1}{8Tm_\chi^2m_S^2K_2(\frac{m_\chi}{T})K_2(\frac{m_S}{T})}\cdot\notag\\
 & 
 \int_{(m_\chi+m_S)^2}^\infty
 \sigma
 \frac{[s-(m_\chi-m_S)^2]}{\sqrt{s}}
 (s-(m_\chi+m_S)^2)
 K_1\left(\frac{\sqrt{s}}{T}\right)
  ds
 \,,
\end{align}
where $K_{1,2}$ are modified Bessel functions of the first and second kind, see, for example, \cite{Cannoni:2013bza}. 
Although we do not need to discretise this collision term in momentum space, we write it in this way to combine it with 
the collision term for decay.
Writing
\begin{align}
\Gamma_{\chi S\rightarrow\psi S} = &\, n_S^{eq} \langle \sigma v \rangle
\end{align}
 and introducing the yield, 
 $Y_\chi = n_\chi/s$, we can write
\begin{align}
\frac{dY_\chi^i}{dx} = - 
\left(\frac{\Gamma_{\chi\rightarrow\psi S}}{\gamma^i} + 
\Gamma_{\chi S\rightarrow\psi S}\right)
(Y^i_\chi - Y^{eq,i}_\chi)\frac{dt}{dx}\,,
\label{eq:Boltzmanndx}
\end{align}
 where $x=m_\chi/T$.  
 
 We will be interested in tracking the relic 
 abundance through changes in the effective number of relativistic degrees of freedom, 
 $g_\ast(T)$, so we use the conservation of 
 entropy to write
 \begin{align}
 \frac{dT}{dt} =&\, \frac{-HT}{1+\frac{T}{3g_\ast(T)}\frac{dg_\ast(T)}{dT}}\,.
 \end{align}
 We then obtain a differential equation for each 
 momentum bin,
\begin{align}
\frac{dY_\chi^i}{dx} = -\frac{
1+{\frac{m_\chi}{3 x g_\ast(T)}\frac{dg\ast(T)}{dT}}}{H x}
\left(\frac{\Gamma_{\chi\rightarrow\psi S}}{\gamma^i} + 
\Gamma_{\chi S\rightarrow\psi S}\right)
(Y^i_\chi - Y^{eq,i}_\chi)\,.
\label{eq:Boltzmanndx}
\end{align}
 Although $\chi$ is at times out of equilibrium we check that 
it never dominates the energy density of the universe, 
so we can use the usual radiation dominated Hubble expansion 
of the universe.
 
\section{Instantaneous Freeze-out}
\label{sec:instantaneous}

With this machinery, we can now turn to exploring various 
mechanisms of dark matter production.  
In this section, we will focus on the region of parameter space 
where the new scalar field exhibits a one-step phase transition, 
\cref{sec:one-step}, and the Yukawa coupling $y_{\chi\psi}$ $\sim10^{-7}$
is large enough to put $\chi$ into equilibrium. We will 
assume $m_\chi \approx m_\psi \approx (30 - 100)  \times m_S(T=0)$.  
This both produces the observed relic abundance of $\chi$ and 
means that at $T \approx \mu_S$ both initial state particles in the 2--to--2 processes 
$\chi \chi \to \psi \psi$, 
$\chi \chi \to S S$ and
$\chi \psi \to S S$ are significantly Boltzmann suppressed, 
dramatically reducing the rate $n \vev{\sigma v}$, so these 
processes can not keep $\chi$ in equilibrium.
We will further assume that $\lambda_p = 10^{-5} - 10^{-3}$ and $y_\psi = 2$.  
These couplings ensure that $S$ and $\psi$ remain in 
equilibrium until the abundance of $\chi$ is set.  
$S$ stays in equilibrium through $SS \leftrightarrow HH$ when it is much 
heavier than the Higgs, and through $SS \leftrightarrow \bar{f}f$, 
where $f$ is a SM fermion, when it is much lighter 
than the Higgs and electroweak symmetry is broken at the temperatures of interest.
For $S$ lighter than the muon, the Yukawa coupling becomes 
too small to keep $S$ in equilibrium, putting a lower limit on the $S$ masses we consider.  
The process $S \leftrightarrow \gamma\gamma$ can also keep $S$ in equilibrium when $S$ has a vev, 
near $T_E$.
The dark sector fermion $\psi$ remains in equilibrium through 
$\psi\psi\leftrightarrow SS$.  

This mechanism, and the others we present in this paper, 
depend heavily on the opening and closing of a kinematic threshold, 
which occurs due to the temperature dependence of the masses 
of $S$ and $\psi$.  As such, we will be interested in the 
region of parameter space where $\chi$ and $\psi$ have a similar 
mass, as is natural if they obtain their masses from the same mechanism.
In this example we will choose 
$m_\chi = 77 \GeV$,
$\tilde{m}_\psi = 74\GeV$.  
We take $y_{\chi\psi} = 5\times 10^{-8}$ for the new Yukawa coupling while 
here and elsewhere we will take 
$y_\chi \approx 0$.  
For the new scalar we will first choose  
$m_S(T=0) = 1\GeV$ and 
$\lambda_S = 1$.  
We can see in \cref{fig:one-step-eff-pot} (right) that with these parameters, 
$\vev{S}(T=0) \approx 1.7\GeV$, so $m_\psi$ will increase from $74\GeV$ at high temperatures 
to $74\GeV + y_\psi \vev{S}(T=0) \approx 77.4\GeV$ at $T\approx0$.  
At high temperatures, $\psi$ is the lightest dark sector particle, while at $T\approx0$, 
$\chi$ takes this role.


\begin{figure}
  \begin{center}
    \begin{tabular}{cc}
      \includegraphics[height=0.45\textwidth]{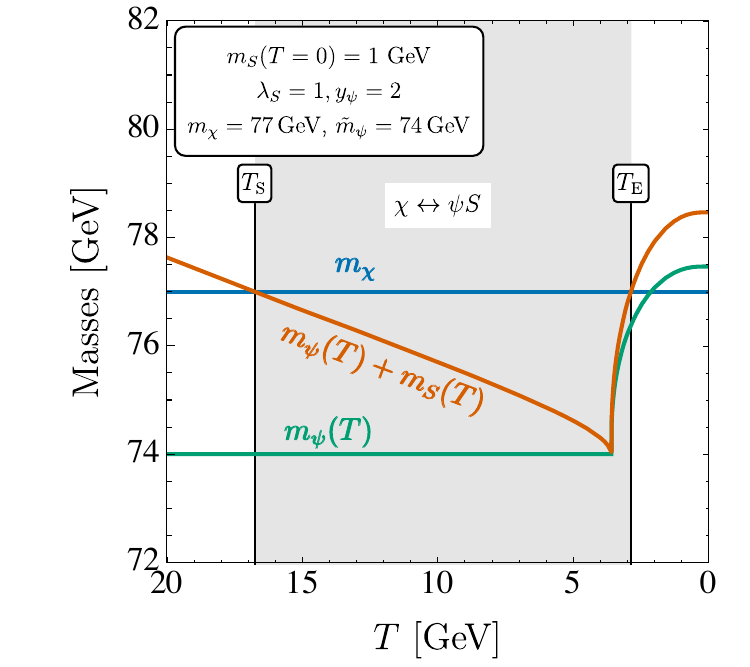} & 
      \includegraphics[height=0.45\textwidth]{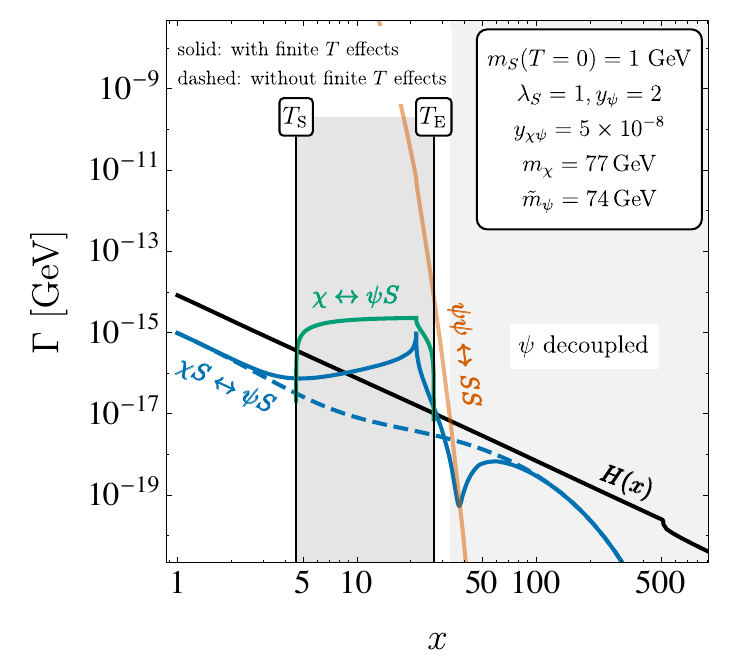}
    \end{tabular}
  \end{center}
  \caption{Left:~the masses of $\chi$ and $\psi$ as a function of 
  temperature, along with the period where the channel 
  $\chi \leftrightarrow \psi S$  is kinematically allowed.  
  Right:~The rates of processes which at some $x$ are larger than the 
  Hubble expansion rate, $H$.  In dashed blue we show the rate of 
   $\chi S \leftrightarrow \psi S$ if finite temperature effects are ignored. 
   In light grey we show the period where $\psi$ is decoupled from the thermal bath.
   }
  \label{fig:sec6-masses-yield}
\end{figure}


The temperature dependent masses for this set of parameters are 
shown in~\cref{fig:sec6-masses-yield} (left).  
We see that at high temperatures, $m_S$ receives 
a large correction and the channel 
$\chi \leftrightarrow \psi S$ is closed.  As the temperature 
reduces, the $T$ dependent corrections become smaller and 
the mass of $S$ reduces.  The channel 
$\chi \leftrightarrow \psi S$
becomes active at $T\approx 17 \GeV$.  This continues 
while $S$ goes through its phase transition, where 
$S$ obtains a vev and its mass starts to increase.  
The vev contributes to $m_\psi$, so both of these effects 
act to close the channel.  
At $T_\text{E} \approx 3\GeV$, 
$m_\chi$ becomes smaller than
$m_\psi(T) + m_S(T)$, so  the process 
 $\chi \leftrightarrow \psi S$ becomes kinematically forbidden.  
 At $T=0$, $m_\chi < m_\psi$, so $\chi$ is the lightest 
 dark sector particle and cannot decay, even via off-shell processes.
 
In~\cref{fig:sec6-masses-yield} (right) we show the rates of 
key processes along with the expansion 
rate of the universe, $H$,  as a function of $x = m_\chi / T$.  When the rate of a process is much larger than 
$H$ then it will effectively keep $\chi$ or $\psi$ in equilibrium, but when it is 
much smaller then it will not.  We see that between $T_\text{S}$ to $T_\text{E}$, 
$\chi \leftrightarrow \psi S$ is the dominant process and will act to put $\chi$ into equilibrium.  
The process $\chi S \leftrightarrow \psi S$ may also have 
a rate larger than the expansion rate of the universe, and will contribute.  
The rate of $\chi S \leftrightarrow \psi S$ shows two resonant features, 
one at $x\sim20$, when $m_S \sim 0$ and the propagator $\psi$ 
can be nearly on-shell, and another at $x\sim40$, 
when $m_\psi \sim m_\chi$ and diagrams (d) and (e) in 
\cref{fig:chi-diagrams} destructively interfere.
Although $\psi$ is 
kept in chemical equilibrium via $\psi \psi \leftrightarrow SS$ until $T_E$, 
it decouples soon after at $x\sim 30$.  Once $\psi$ has departed from chemical 
equilibrium, the relic abundance of  
$\chi$ cannot be reduced, since any $\psi$ produced will simply decay back to $\chi$.
We also show the rate of  $\chi S \leftrightarrow \psi S$ if finite temperature 
effects are ignored.  We see that in this case the $\chi S \leftrightarrow \psi S$ 
rate is always smaller than the Hubble rate, 
so $\chi$ will freeze-out while relativistic.
All other processes involving $\chi$ have rates orders of magnitude smaller 
than the Hubble rate.

\begin{figure}
  \begin{center}
    \begin{tabular}{cc}
      \includegraphics[height=0.45\textwidth]{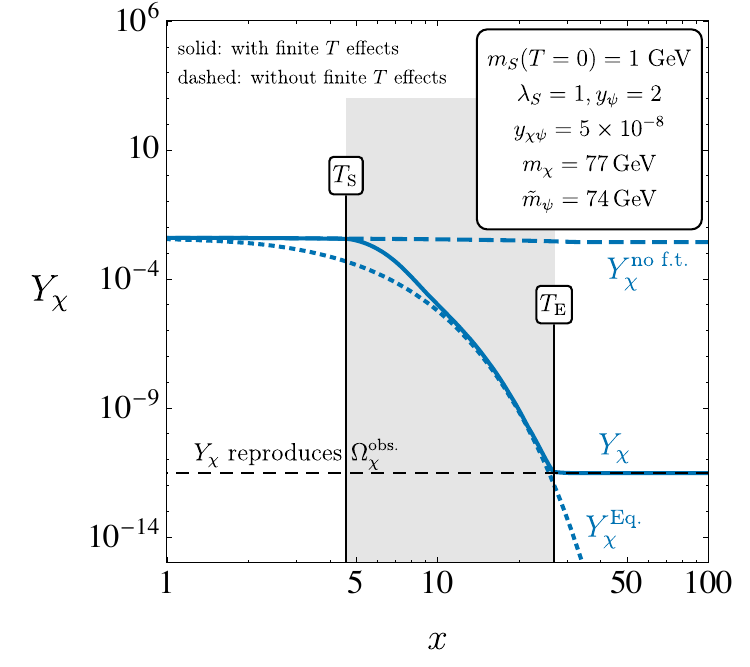} &
      \includegraphics[height=0.45\textwidth]{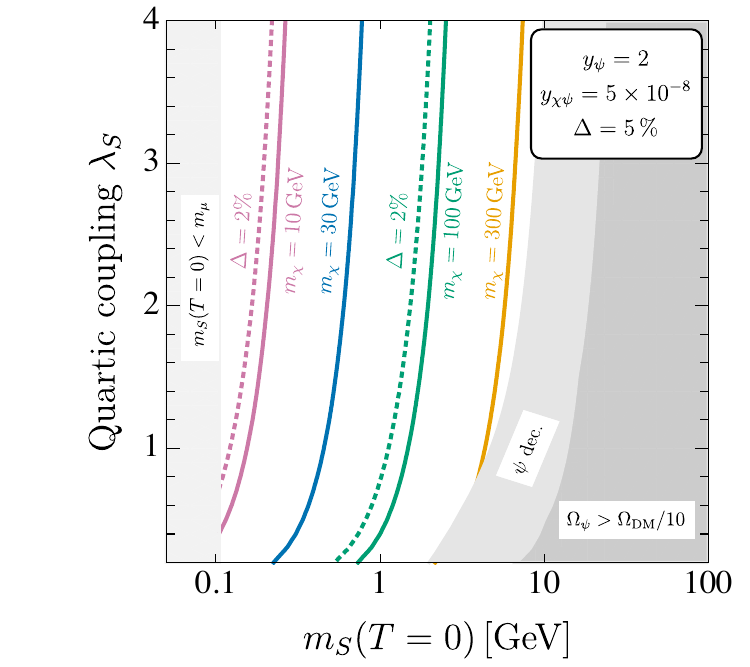}
    \end{tabular}
  \end{center}
  \caption{
  Left:~the evolution of the yield of $\chi$ as a function of 
  temperature (solid) with the equilibrium yield of $\chi$ (dotted) 
  and the yield calculated without 
  accounting for the finite temperature effects (dashed). 
  Also shown is  the yield which is required to obtain the observed relic abundance 
  of $\chi$ (black, dashed).
  Right:~the required values of $m_S(T=0)$ and $\lambda_S$ for various $\chi$ masses, for $\Delta=5\%$ and $2\%$.
  In the left light-grey region $S$ is not in equilibrium throughout the decay of $\chi$, 
  in the central grey region $\psi$ decouples before $T_\text{E}$, while 
  in the right dark-grey region $\psi$ is not a subdominant relic.
   }
  \label{fig:sec6-param-space}
\end{figure}

In~\cref{fig:sec6-param-space} (left) we show the evolution of the 
yield of $\chi$.  
At small $x$ there are no processes connecting $\chi$ to the thermal bath 
with a rate larger than the Hubble rate, so $\chi$ freezes-out.  
Then, at $T_S$ the rate of $\chi \leftrightarrow \psi S$ (and later $\chi S \leftrightarrow \psi S$) 
becomes larger than the Hubble rate and the $\chi$ yield comes 
back into equilibrium.
At $T_\text{E}$, however, these channel suddenly becomes inefficient and, 
since there are no longer any processes keeping $\chi$ in 
equilibrium, it instantaneously freezes-out. 
We see that in this scenario, the relic abundance of $\chi$ is 
set not by usual, smooth freeze-out, resulting from an interplay 
between a slowly varying annihilation or decay rate and the Hubble rate, 
but by a sudden and dramatic change in rates.  
If these finite temperature effects are ignored, the final abundance is overestimated 
by $\approx 10$ orders of magnitude, since the rate of the most relevant process 
$\chi S \leftrightarrow \psi S$ remains below the Hubble rate throughout.

This means that the resulting yield is not a function of 
the coupling constant (provided the assumptions above are satisfied), 
but only of the mass of the dark matter candidate and $T_\text{E}$.  
The final yield is simply given by
\begin{align}
  Y_\chi^\infty =&\, Y_\chi^\text{Eq.}(T=T_\text{E})\,,
  \label{eq:sec6-yield}
\end{align}
which is only a function of $m_\chi$ and $T_\text{E}$ (which 
in turn is a function of $\tilde{m}_\psi$, $y_\psi$ and the 
effective potential).


We now consider the parameter space of this mechanism.  
Rather than considering the full 7-dimensional parameter space, 
we explore the impact of varying one or two parameters at a time.  
The final relic abundance is insensitive to $y_{\chi\psi}$ as 
long as $\chi \leftrightarrow \psi S$ is fast enough to keep 
$\chi$ in equilibrium (providing a lower bound on $y_{\chi\psi}$) 
and as long as $\chi S \leftrightarrow \psi S$ does not significantly 
deplete $\chi$ after $T_\text{E}$ (providing an upper bound on 
$y_{\chi\psi}$).  For the benchmark values of the other parameters, 
this is satisfied for $3\times 10^{-8} < y_{\chi\psi} < 1\times 10^{-7}$.
As discussed above, the mechanism occurs while 
$y_\psi$ is large enough to keep 
$\psi$ in equilibrium while the processes $\chi \leftrightarrow \psi S$ and $\chi S \leftrightarrow \psi S$ 
occur.  
If $\psi$ goes out of equilibrium before $T_\text{E}$, 
the processes will not reduce the energy density in the 
dark sector and, once $m_\chi < m_\psi$, $\psi$ will decay to $\chi$.
In this variation of the mechanism, the relic abundance is set 
by the freeze-out of $\psi$, not via the dramatic change in the 
rates of $\chi \leftrightarrow \psi S$ and $\chi S \leftrightarrow \psi S$ at $T_\text{E}$.
However, in both scenarios, the Yukawa coupling $y_\chi$ should be small enough that 
the 2--to--2 process $\chi \chi \leftrightarrow S S$ is inefficient by $T_\text{E}$ 
or when $\psi$ freezes-out.

In \cref{fig:sec6-param-space} (right) we show the mass of $\chi$ required to 
produce the observed relic abundance, for $\Delta = 5\%$ and $2\%$.  We see that 
$m_\chi$ is $30 -100$ times larger than $m_S(T=0)$, depending on $\lambda_S$.  
Although $m_S(T=0)$ depends 
on $\mu_S$, the $S$ vev also depends inversely on the square root of $\lambda_S$.   
This means that as $\lambda_S$ becomes smaller, the $S$ vev gets larger.  The temperature 
of the phase transition, which is close to $T_\text{E}$, also increases.  These two effects mean that $m_\chi$ has to 
increase to compensate.  
Reducing the mass splitting $\Delta$ has the effect of requiring a larger $m_\chi$.  
This is because a smaller $\Delta$ means that the important 1-to-2 process 
closes at a higher $T$, so $Y_\chi^\text{Eq.}(T=T_\text{E})$ becomes larger.  
To yield the observed relic abundance, $m_\chi$ then has to increase.
If the mass splitting is much larger than $5\%$ then the
$\vev{S}$ contribution to $m_\psi$ at $T=0$ is not large enough to close the mass gap, so 
$m_\chi > m_\psi(T=0)$.
As mentioned above, if $S$ is lighter than the muon then 
it is not in equilibrium throughout the process, so the decay does not reduce the 
energy density of the dark sector.  
For $m_\mu < m_S(T=0) \lesssim 1\GeV$, either $\lambda_p$ must be reduced below $\lambda_p = 10^{-3}$ or there is some 
tuning between the $\mu_S^2$ term and $\lambda_p\vev{h}S^2/4$ to achieve this low mass.  
If $\lambda_p$ is reduced, care must be taken 
that $S$ remains in equilibrium throughout.  We find that we require $\lambda_p \gtrsim 10^{-5}$ 
in this low $m_S$ region of parameter space.
On the other hand, as $m_S(T=0)$ increases, $m_\psi$ also increases (along with $m_\chi$).  
For the parameters chosen, when $\psi$ is heavy it can freeze-out (at $T=T_\psi^\text{f.o.}$) 
at temperatures higher than $T_\text{E}$.  
In this case, dark sector fermion number conservation 
implies that the 
$\chi$ abundance is given by 
\begin{align}
  Y_\chi^\infty =&\, Y_\chi^\text{Eq.}(T=T_\psi^\text{f.o.})\,.
  \label{eq:sec6-yield-2}
\end{align}
With $y_\psi = 2$,  if $m_\psi \gtrsim 1\TeV$ then $\psi$ will freeze-out with a relic abundance greater 
than 10\% of the observed relic abundance, so the mechanism we are discussing 
will not be the dominant mechanism in setting the relic abundance.  
The quartic coupling $\lambda_S$ can become larger than 4 
but at some point it becomes non-perturbative and the one-loop analysis 
breaks down.

\section{Decaying Dark Matter with a One-Step Phase Transition}
\label{sec:one-step-decay}

Now we continue to consider the region of parameter space 
where the new scalar field exhibits a one-step phase transition, 
discussed in \cref{sec:one-step}, but assume a smaller Yukawa coupling, 
$y_{\chi\psi} \lesssim10^{-7}$.  With a coupling this small, 
$\chi$ either freezes-out while relativistic or never comes 
into equilibrium with the thermal bath (as in the freeze-in 
scenario~\cite{Baker:2017zwx}).  Here we will assume that there is 
some UV physics which connected $\chi$ to the thermal 
bath after reheating, so that $\chi$ was in equilibrium at high temperatures.  
In this scenario $\chi$ will have frozen out when relativistic, then 
$\chi \leftrightarrow \psi S$, and to a lesser extent $\chi S \leftrightarrow \psi S$, 
bring $\chi$ towards equilibrium 
after $T_S$.  This means that 
$\chi$ will deplete and its abundance will approach 
the equilibrium abundance. 
At $T_\text{E}$ these process become inefficient and 
the yield of $\chi$ stabilises.  
We will require $m_\chi \approx m_\psi \approx 30 \, m_S$ 
to obtain the observed relic abundance.   
With these masses and couplings 
 the rates of  $\chi \leftrightarrow \psi S$ and $\chi S \leftrightarrow \psi S$ are 
 the only processes to ever have a rate greater than the Hubble rate.  
 As discussed in \cref{sec:instantaneous}, $\psi$ and 
 $S$ remain in equilibrium throughout the decay process.

\begin{figure}
  \begin{center}
    \begin{tabular}{cc}
      \includegraphics[height=0.45\textwidth]{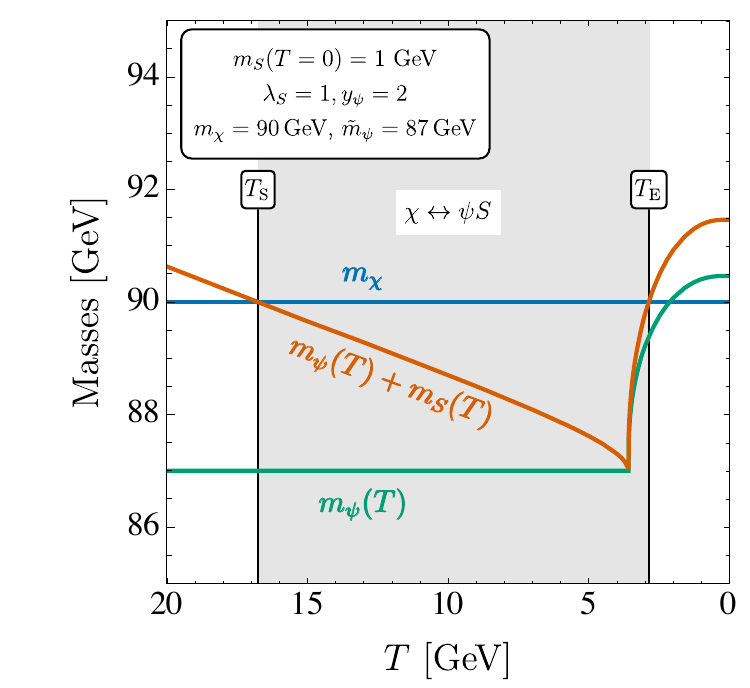} & 
      \includegraphics[height=0.45\textwidth]{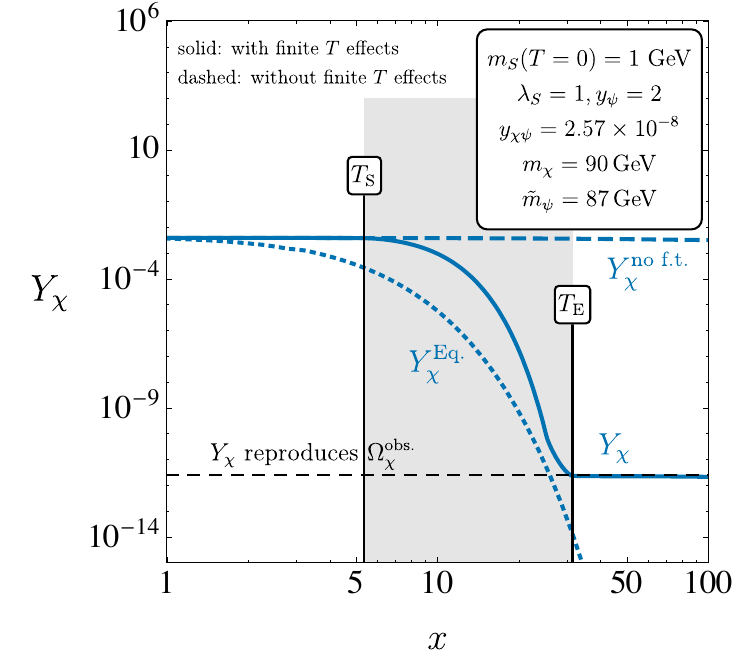}
    \end{tabular}
  \end{center}
  \caption{
  Left:~the masses of $\chi$ and $\psi$ as a function of 
  temperature, along with the period where the channel 
  $\chi \leftrightarrow \psi S$  is kinematically allowed.  
  Right:~the evolution of the yield of $\chi$ as a function of 
  temperature (solid) with the equilibrium yield of $\chi$ (dotted) 
  and the yield calculated without 
  accounting for the finite temperature effects (dashed). 
  Also shown is  the yield which is required to obtain the observed relic abundance 
  of $\chi$ (black, dashed).
   }
  \label{fig:sec7-masses-yield}
\end{figure}

To describe the mechanism in this case, we will keep
the same scalar parameters chosen above but slightly increase 
the dark sector fermion masses (we take in this example 
$m_\chi = 90\GeV$ and $\tilde{m}_\psi = 87\GeV$) and 
reduce the Yukawa coupling to  
$y_{\chi\psi} \lesssim10^{-7}$ (here we take 
$y_{\chi\psi} = 2.57\times10^{-8}$).  
The temperature 
dependence of the masses is shown in \cref{fig:sec7-masses-yield} (left).
We see that with these parameter choices, 
\cref{fig:sec6-masses-yield} (left) is simply shifted to slightly higher masses and the channel 
$\chi \leftrightarrow \psi S$ is open in the same temperature range, 
$17 \GeV \gtrsim T \gtrsim 3 \GeV$.  


In  \cref{fig:sec7-masses-yield} (right) we show the evolution of the 
yield of $\chi$ with $x$.  As expected for $y_{\chi\psi} \lesssim 10^{-7}$, 
$\chi$ freezes-out while it is still relativistic.  However, at $T_\text{S} \approx 17\GeV$ 
the channel $\chi \leftrightarrow \psi S$ becomes kinematically allowed and 
the $\chi$ abundance reduces, approaching the equilibrium curve.  
The $\chi S \leftrightarrow \psi S$ channel also contributes to a lesser extent.  
However, in this case, the rate is 
not fast enough to reach equilibrium and $Y_\chi$ remains larger than 
$Y_\chi^\text{Eq.}$ at all times.  This continues until the channel closes at 
$T_\text{E} \approx 3\GeV$ and the yield of $\chi$ becomes fixed until the present day, 
reproducing the observed relic abundance.

In this case the final yield is not set by $Y_\chi^\text{Eq.}$ at $T_\text{E}$, but 
by the amount of $\chi$ that has decayed in the period from 
$T_\text{S}$ to $T_\text{E}$.  This depends on $y_{\chi\psi}$ and on the 
temperature dependent masses of the particles.  As such there is no simple 
formula for calculating the resulting abundance; the final abundance 
must be calculated numerically.


\begin{figure}
  \begin{center}
    \begin{tabular}{cc}
      \includegraphics[height=0.45\textwidth]{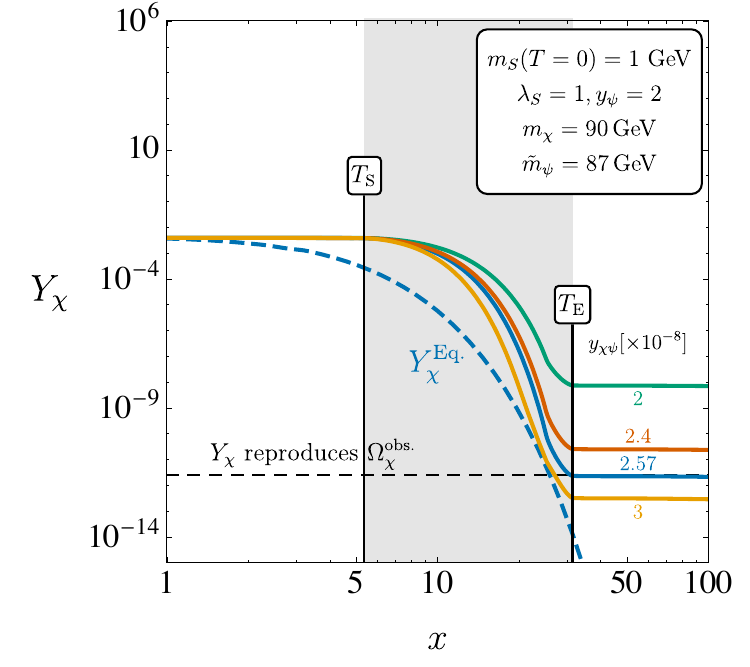} &
      \includegraphics[height=0.45\textwidth]{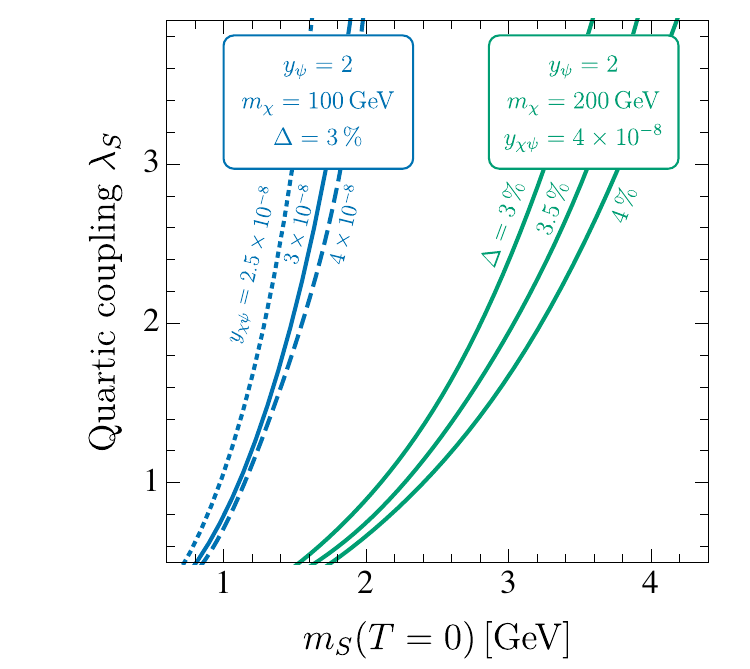}
    \end{tabular}
  \end{center}
  \caption{
  Left:~the evolution of the yield for different values of $y_\chi$ (solid lines). 
  The equilibrium abundance of $\chi$ is dashed blue while the 
  yield required to reproduce the observed DM relic abundance is 
  black dashed.   
  Right:~the $m_S(T=0)$ and $\lambda_{S4}$ which reproduce 
  the observed DM relic abundance for different parameter choices.  
  For the dotted line $\psi$ freezes-out before $T_\text{E}$, while for the dashed line 
  $\chi$ returns to equilibrium.
   }
  \label{fig:sec7-param-space}
\end{figure}


We now vary the parameters of the model to see their dependence.  
We first point out that the relic abundance obtained is exponentially 
sensitive to the coupling $y_{\chi\psi}$.  
In \cref{fig:sec7-param-space} (left) we show the evolution of 
the yield for different values of $y_{\chi\psi}$.  
As seen above, for a coupling of $y_{\chi\psi} = 2.57\times10^{-8}$ we obtain 
$\Omega_\chi h^2 = 0.12$.  
With a smaller coupling of 
$y_{\chi\psi} = 2\times10^{-8}$ we obtain $\Omega_\chi h^2 \sim 340$, 
$y_{\chi\psi} = 2.4\times10^{-8}$ gives $\Omega_\chi h^2 = 1.2$,
while  $y_{\chi\psi} = 3\times10^{-8}$ gives $\Omega_\chi h^2 = 0.015$.  
We see that a $\approx 10\%$ 
change in the coupling changes the final abundance by an order of 
magnitude.  As such, this mechanism provides no answer to the 
coincidence problem (that $\Omega_\chi \approx 5 \, \Omega_\text{SM}$).

In \cref{fig:sec7-param-space} (right) we vary the effective potential, $m_\chi$, 
$y_{\chi\psi}$ and $\Delta$, and plot the curve where the observed relic abundance 
is reproduced.  
We see that the observed abundance is reproduced for $m_\chi \approx (50 - 200) \times m_S(T=0)$.  
As in \cref{sec:instantaneous}, a smaller quartic 
coupling $\lambda_S$ results in a larger $S$ vev and phase transition temperature, 
resulting in a heavier $\chi$.  
Increasing the mass of $S$ similarly provides a larger vev and a larger 
phase transition temperature, requiring a heavier $\chi$.  
A larger $y_{\chi\psi}$ means that $\chi$ depletes faster so $T_\text{E}$ needs to occur at a higher 
temperature, which is achieved with a heavier $S$ or a smaller $\lambda_S$.  
Similarly, a larger $\Delta$ means that $T_\text{E}$ is naturally lower, so this needs 
to be compensated with a larger $S$ mass or a smaller $\lambda_S$.  
When $S$ is lighter than the muon, $S$ must couple via a first generation SM Yukawa coupling, 
which is so small that $S$ does not stay in equilibrium throughout the process.  
The upper bound on $m_S$ is again determined by requiring that $\psi$ freezes-out as a 
 subdominant relic, around $m_\psi \approx 1 \TeV$, 
but we show just a small region of the 
possible parameter space to demonstrate the impact of varying $y_{\chi\psi}$ and $\Delta$. 
It may also happen, as in \cref{sec:instantaneous}, that $\psi$ freezes-out before $T_\text{E}$.  This occurs for 
the dotted line ($y_{\chi\psi} = 2.5\times10^{-8}$), since the lower $m_S(T=0)$ reduces $T_\text{E}$ 
below the $\psi$ decoupling temperature.
For the dashed line ($y_{\chi\psi} = 4\times10^{-8}$) the coupling is large enough that 
$\chi$ returns to equilibrium, so the situation is as discussed in \cref{sec:instantaneous}.
The quartic coupling $\lambda_S$ can also in principle be increased to the perturbativity limit.

\section{Decaying Dark Matter with a Two-Step Phase Transition}
\label{sec:two-step-decay}

Finally we consider the region of parameter space 
where $\lambda_p$ is not negligible, but 
$\mathcal{O}(1)$.  As discussed in 
\cref{sec:two-step}, this can lead to a two-step phase transition 
where first the new scalar $S$ obtains a vev and later, at the onset of 
electroweak symmetry breaking (EWSB), the SM Higgs obtains a vev and $\vev{S}$ goes to zero. 
This two-step phase transition only occurs in the region 
of parameter space where $\mu_S \approx \mu_H$, which means 
that $m_S(T=0) = \mathcal{O}(100\GeV)$.  
To maintain $m_\chi \gtrsim 30\,m_S(T=0)$,
so that the abundance is set by decay, 
we require $m_\chi, m_\psi  \approx 4$ -- $5\TeV$.
Since $\psi$ is now quite heavy, it will naturally 
freeze-out with a large relic abundance.  To counter 
this and to ensure that $\psi$ is a subdominant relic, 
we have to take $|y_\psi| \gtrsim 4$ which, while large, 
is still perturbative.  However, corrections from higher orders 
in the perturbative expansion will not be as small as usually assumed.  
With this two-step phase transition, we will take 
$y_\psi < 0$, so that $\psi$ becomes lighter when 
$S$ obtains a vev.
This allows the channel $\chi \leftrightarrow \psi S$ 
to open when $S$ obtains a vev, and to close when the
$S$ vev goes to zero. 
This situation can lead to the observed relic abundance either by 
instantaneous freeze-out when $y_{\chi\psi}$ 
is $\sim 10^{-7}$, as in 
\cref{sec:instantaneous}, or with a period of DM decay 
with $y_{\chi\psi} \lesssim10^{-7}$, as in 
\cref{sec:one-step-decay}.  Here we choose to explore the 
latter, to make a connection with the mechanism described 
in~\cite{Baker:2016xzo}.  

In~\cite{Baker:2016xzo} it was shown that a new $su(2)_L$ triplet scalar $S_3$ 
offers decay channels to a singlet DM 
candidate when there are two extra dark sector $su(2)_L$ triplet fermions, 
$\psi_3$ and $\psi_3'$.  
In this case, there are the channels $\chi \leftrightarrow \psi_3^+ W^-$ 
and $\chi \leftrightarrow \psi_3'^+ W^-$ in addition to the process discussed in the present work, 
$\chi \leftrightarrow \psi_3 S_3$.  
However, as we demonstrate 
here, this extra structure is not required and the observed relic abundance 
can be obtained through the $\chi \leftrightarrow \psi S$ process.


\begin{figure}
  \begin{center}
    \begin{tabular}{cc}
      \includegraphics[height=0.45\textwidth]{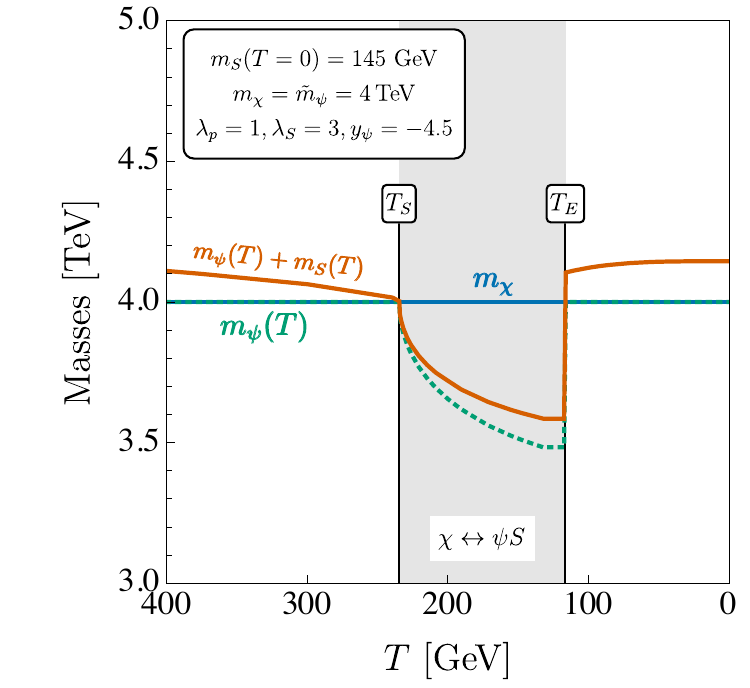} & 
      \includegraphics[height=0.45\textwidth]{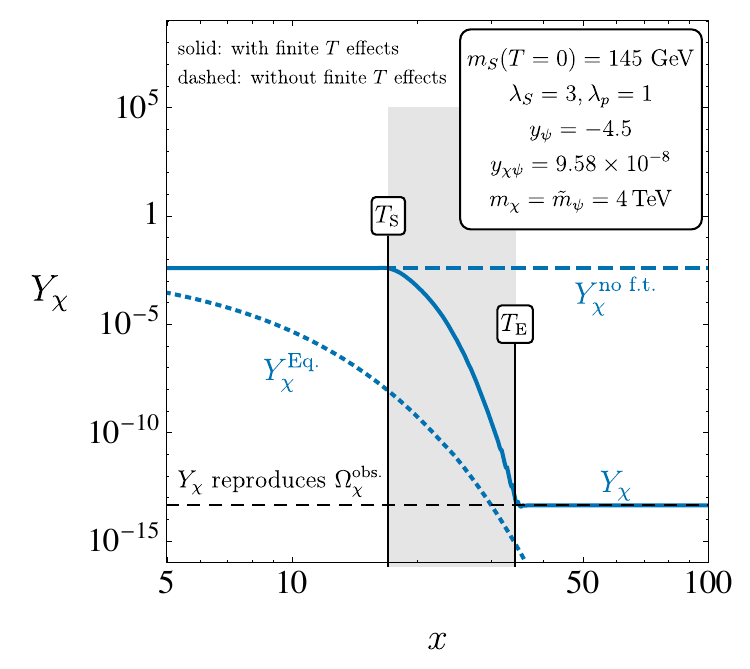} 
    \end{tabular}
  \end{center}
  \caption{
  Left:~the masses of $\chi$ and $\psi$ as a function of 
  temperature, along with the period where the channel 
  $\chi \leftrightarrow \psi S$  is kinematically allowed. 
  Right:~the evolution of the yield of $\chi$ as a function of 
  temperature (solid) with the equilibrium yield of $\chi$ (dashed) 
  and the yield which is required to obtain the observed relic abundance 
  of $\chi$ (black, dashed). 
   }
  \label{fig:sec8-masses-yield}
\end{figure}


In \cref{fig:sec8-masses-yield} we again show (left) the 
masses as a function of $T$.  
Recall that $m_\psi(T) = \tilde{m}_\psi + y_\psi \vev{S}$.  
We see that in this case  
$m_\chi = m_\psi = 4\TeV$, as long as $\vev{S} = 0$, c.f., 
\cref{fig:two-step-masses-vevs}.  
However, when $\vev{S} \neq 0$, there is a large negative 
contribution to $m_\psi$, since $y_\psi < 0$.  The mass of 
$S$ also becomes relatively small between $T_\text{S}$ and $T_\text{E}$, 
so the channel $\chi \leftrightarrow \psi S$ opens.  At $T_\text{E}$ there 
is a first-order phase transition and, since $\vev{S}$ goes to 
zero, $m_\psi(T)$ returns to $\tilde{m}_\psi$ and 
the channel $\chi \leftrightarrow \psi S$ abruptly closes.  

As mentioned above, the large Yukawa coupling $|y_\psi| \gtrsim 4$ 
ensures that $\psi$ remains in equilibrium throughout the depletion of $\chi$. 
$S$ has a large portal coupling so can easily stay in equilibrium via 
$SS\leftrightarrow HH$.  
As discussed in \cref{sec:thermal-bath}, the rates of all 2--to--2
processes involving $\chi$, except $\chi S \leftrightarrow \psi S$, 
remain significantly below the Hubble rate at all times.  
The rate of $\chi S \leftrightarrow \psi S$ is suppressed when 
$m_\chi \approx m_\psi(T)$ but enhanced when $m_S(T) \approx 0$.  
This means that the rate of $\chi S \leftrightarrow \psi S$ can be larger than the Hubble 
rate in the period where $S$ obtains a vev.  Although it remains 
subdominant to $\chi \leftrightarrow \psi S$ in this period, we include it 
in out numerical calculations.  Like $\chi \leftrightarrow \psi S$, the rate of  $\chi S \leftrightarrow \psi S$ 
abruptly reduces below the Hubble rate at $T_\text{E}$.

In \cref{fig:sec8-masses-yield} (right) we see the yield of 
$\chi$ as a function of $x$.  As in \cref{sec:one-step-decay}, 
$\chi$ initially freezes-out when relativistic, before beginning to deplete 
at $T_\text{S}$.  The depletion 
continues until it abruptly halts at $T_\text{E}$.  It stops more abruptly 
than in \cref{sec:one-step-decay} due to the first-order 
phase transition, which quickly changes the mass of $\psi$ 
at $T_\text{E}$.  After $T_\text{E}$, $\chi$ stabilises and the yield 
remains constant.  We see that 
there is a choice of $y_\psi$ such that the observed 
relic abundance is obtained.  Again, once $\chi$ has 
departed from equilibrium at high temperatures, it does not 
return to equilibrium. 
We see that properly accounting for these finite temperature effects is essential 
for calculating the correct abundance.


\begin{figure}
  \begin{center}
    \begin{tabular}{cc}
      \includegraphics[height=0.45\textwidth]{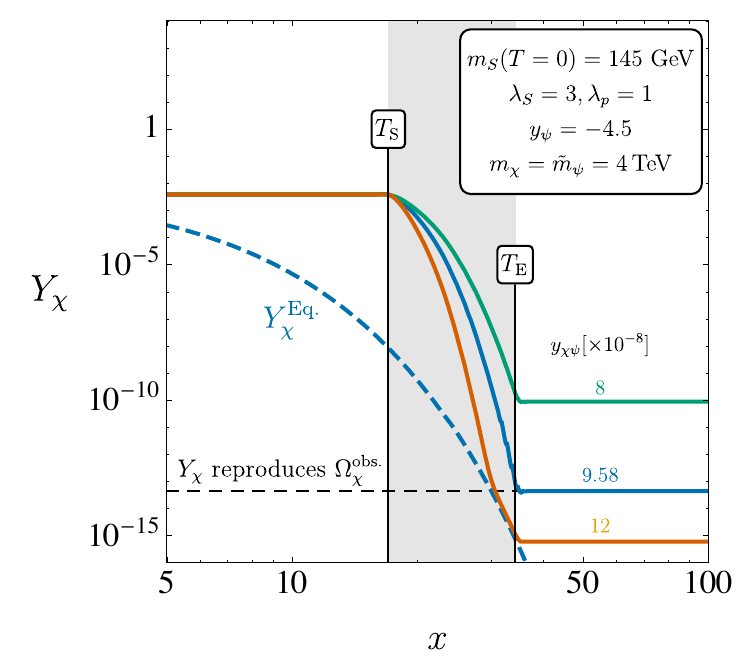} &
      \includegraphics[height=0.45\textwidth]{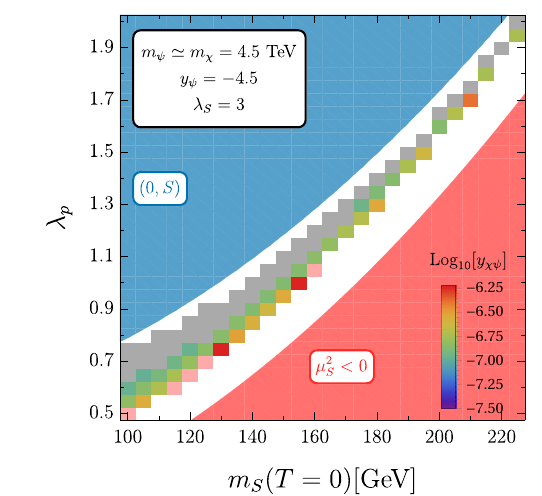}
    \end{tabular}
  \end{center}
  \caption{
  Left:~the evolution of the yield for different values of $y_\chi$.  
  Right:~the $y_\chi$ required to produced the observed relic abundance in the 
  $\lambda_S$ -- $m_S(T=0)$ plane.  In the red region $S$ does not obtain a vev, 
  while in the blue region electroweak symmetry is not broken at $T=0$. 
  The grey pixels show where $\psi$ freezes-out before $T_\text{E}$ while 
  at the pink pixels the equilibrium abundance of $\chi$ at $T_\text{E}$ 
  is too large for $Y_\chi$ to reduce to the observed value.
   }
  \label{fig:sec8-param-space}
\end{figure}


We now consider the parameter space of this scenario.  In 
\cref{fig:sec8-param-space} (left) we show the evolution of the yield 
for different Yukawa couplings $y_{\chi\psi}$.  We see that the final 
yield is again exponentially sensitive to this coupling.  
A small change in the coupling 
results in an orders of magnitude change in the final yield.  
When the Yukawa coupling is $y_{\chi\psi} \gtrsim 1.2\times10^{-7}$ 
we see that the rate is fast enough to bring $\chi$ back into equilibrium.  
We here explicitly see the relationship between instantaneous freeze-out 
and decaying dark matter.  If $\chi$ does go back into equilibrium, 
the final abundance is simply set by the equilibrium abundance at $T_\text{E}$, 
as in \cref{sec:instantaneous}.  We note that the abundance of 
$\chi$ cannot go below the equilibrium abundance with this mechanism.

In \cref{fig:sec8-param-space} (right) we show the value of $y_{\chi\psi}$ 
required to produce the equilibrium abundance of $\chi$ in the 
$m_S(T=0)$ -- $\lambda_p$ plane.  
We have increased the $\chi$ and $\psi$ mass to 4.5 TeV, since 
otherwise the equilibrium abundance of $\chi$ is above the observed 
abundance at $T_\text{E}$ for many points, meaning that decay cannot set the abundance 
(this is still the case for some points, shown in pink). 
In the red region, 
$\mu_S^2$ is negative so $S$ never obtains a vev and $\chi$ does not decay.  
In the blue region, the deepest minima at zero temperature have 
$\vev{S} \neq 0$ and $\vev{h} = 0$, so electroweak symmetry 
is not broken (i.e., this region is unphysical).  In the region between these 
curves we show a pixellated region where the universe successfully 
transitions to the EWSB minimum.  We see that larger couplings $y_{\chi\psi}$  
are required near the red region, where symmetry breaking is weak and 
$S$ only obtains a vev for a short amount of time.  Closer to the blue region  
the required Yukawa coupling is smaller, as there is a longer amount of time 
in which $\chi$ can decay. 
The grey pixels show where $\psi$ freezes-out before $T_\text{E}$. 

We do not show the variation with $m_\chi$ and $m_\psi$ since this mechanism 
only works in a relatively small window.  If these masses become much larger, then 
$\psi$ is no longer a subdominant relic.  On the other hand, since $T_\text{E} \approx 100 \GeV$, 
if the fermions are much lighter, then the equilibrium abundance at $T_\text{E}$ 
is too large and $\chi$ overcloses the universe.

\section{Experimental Constraints}
\label{sec:exp-const}

We now briefly consider experimental tests of the model 
in different regions of its parameter space.  
The usual probes of dark matter, direct and indirect detection 
of the galactic population and direct dark matter production at colliders, 
may be effective for the parameter space 
for instantaneous freeze-out,
considered in \cref{sec:instantaneous}. 
However, since in this section the relic abundance is not determined 
by $y_{\chi\psi}$, there is no expectation of a coupling of a certain 
size (as there is in typical freeze-out scenarios).  
In \cref{sec:one-step-decay,sec:two-step-decay},
where we consider decaying dark matter,  
these searches are hindered by the small coupling $y_{\chi\psi}$, as is common for 
freeze-in models.  However, direct detection is beginning to probe even these small 
couplings~\cite{Hambye:2018dpi}.

Aside from directly searching for $\chi$, the model may first be probed via 
its other particle content.  
In all cases the energy density in $\chi$ must be passed to 
the SM thermal bath.  This means that $S$ and $\psi$ must be in 
equilibrium while the $\chi \leftrightarrow \psi S$ channel 
is open, which in turn 
means that the $S^2 |H|^2$ portal coupling must be $\approx 10^{-3}$ (although for masses less than a GeV 
it could be as small as $10^{-5}$). 
If $S$ is lighter than $m_h/2$, it may be detected by precision measurements of the Higgs.  
The current limit excludes $\lambda_p \gtrsim 10^{-2}$~\cite{Aad:2015pla,Khachatryan:2016whc} while future colliders 
such as the ILC and CEPC will probe $\lambda_p \gtrsim 10^{-3}$~\cite{Chacko:2013lna,Ko:2016xwd,Han:2016gyy}. 
Pairs of heavy $S$ particles may be produced via an off-shell Higgs at proton colliders.  
A future $100\TeV$ machine with $30\text{ ab}^{-1}$ of integrated luminosity 
will be able to exclude the two-step phase transition region of \cref{sec:two-step-decay} 
at  $2\sigma$ 
but not discover an $S$ in this region at  $5\sigma$~\cite{Curtin:2014jma,Craig:2014lda,Chen:2017qcz}. 
It will not be able to probe 
the scalar sector in \cref{sec:instantaneous,sec:one-step-decay}. 

In \cref{sec:instantaneous,sec:one-step-decay}, any remaining 
$\psi$ and $S$ can decay, so these will not make any 
of the dark matter relic abundance.  However, in
\cref{sec:two-step-decay} there is a subdominant population of both $\psi$ 
and $S$ ($S$ has no vev at $T=0$ so the $\mathbb{Z}_2$ symmetry may be intact).  
Assuming that the $\mathbb{Z}_2$ symmetry is not 
softly broken by effects from particles beyond this simple model, 
the subdominant populations of $S$ and $\psi$ may be detected 
by direct or indirect detection experiments.  However, the rates seen in these 
experiments depend on their abundances, which are unconnected to the 
$\chi$ abundance.

In the scenario in \cref{sec:two-step-decay} 
there is a first-order phase transition, as $h$ obtains a vev and $S$ loses 
its vev.  
If this phase transition is strong enough there is the interesting possibility of detecting 
a stochastic gravitational wave background resulting from 
the first order phase transition in upcoming space based interferometers 
such as LISA~\cite{Caprini:2015zlo}.

\section{Conclusions}

In this work we have explored the impact of finite temperature corrections 
on the relic abundance of a simple model of dark matter, which consists of 
two dark sector fermions (one being a dark matter candidate, $\chi$, and 
a further dark sector fermion $\psi$) and 
a new scalar, $S$.  
We calculate the leading finite temperature effects, using the effective 
potential of the new scalar and the SM Higgs, and track 
the dark matter abundance using Boltzmann equations.  
We emphasise that in different regions of parameter space the 
new scalar can either simply obtain a vev (which we focus on in \cref{sec:instantaneous,sec:one-step-decay}), 
or pass through a two-step phase transition 
during EWSB (\cref{sec:two-step-decay}).
These finite temperature effects provide corrections to the scalar and 
fermion masses.
We assume that $\chi$ was in equilibrium at high temperatures and 
focus on regions of parameter space where decays and 
inverse decays $\chi \leftrightarrow \psi S$ are kinematically 
allowed for a period of time. 
We also ensure that $S$ and $\psi$ remain in 
equilibrium with the SM thermal bath during the time of interest.
We have explored the parameter space of the model and 
identified different regions where finite temperature effects lead to 
non-standard mechanisms of dark matter production.

In \cref{sec:instantaneous} we discussed instantaneous freeze-out and considered the parameter space where 
$\chi$ can quickly come into equilibrium through the processes  
$\chi \leftrightarrow \psi S$ and $\chi S \leftrightarrow \psi S$.  
When these channels abruptly become inefficient, $\chi$ instantaneously freezes-out, 
resulting in a tight relationship between the temperature of decoupling 
and the resulting relic abundance.  We find viable dark matter candidates 
in the range $m_\chi \approx (3 \GeV - 300 \GeV)$, although $\chi$ and 
$\psi$ should have similar mass (at the 5\% level) to avoid significant fine tuning.  
However, the final abundance depends exponentially on the temperature at which 
the $\chi \leftrightarrow \psi S$ and $\chi S \leftrightarrow \psi S$ channels become inefficient.

In \cref{sec:one-step-decay} we 
discuss the case where $\chi$ is 
only weakly coupled to the thermal bath and 
consider decaying dark matter via a one-step phase transition.  
In this case, rather than quickly 
coming into equilibrium, 
the yield of $\chi$ slowly approaches the equilibrium yield.  
The final abundance of $\chi$ depends on all of the model parameters in 
a complicated manner.  We again find viable dark matter candidates 
in the range $m_\chi \approx (3 \GeV - 300 \GeV)$.  In this scenario, 
the final abundance is exponentially sensitive to the $\bar{\psi}\chi S$ coupling.

In \cref{sec:two-step-decay} we again focus on the case where 
$\chi$ is weakly coupled and
describe decaying dark matter via a two-step phase transition  
(although we point out that instantaneous freeze-out 
can also occur during a two-step phase transition).  
We show that by changing the sign of a Yukawa coupling, we can again 
achieve the observed dark matter abundance with a period of decay.  
In this case $S$ must have a mass similar to the SM Higgs, resulting in a 
smaller region of viable parameter space.

Finally we survey the experimental prospects for detecting 
particles in these scenarios.  
In all cases the energy density in $\chi$ must be passed to 
the SM thermal bath.  This means that $S$ must be in 
equilibrium while the dominant $\chi \leftrightarrow \psi S$ channel 
is open.  The $S^2 |H|^2$ portal coupling cannot be too small, 
so a light $S$ may be detected by precision measurements of the Higgs, while 
a heavy $S$ may be detected at a future $100\TeV$ proton collider.   
Direct and indirect detection of $\chi$ is usually hindered by its 
small couplings.
In \cref{sec:two-step-decay} the subdominant population of $\psi$ 
may be detected by these means.  
In this case there are also interesting possibilities in detecting 
a stochastic gravitational wave background resulting from 
a first order phase transition.

Although we study a simple model of dark matter, these mechanisms 
for setting the dark matter relic abundance, some which we discuss here for the first time, 
may be relevant in a 
wide range of Beyond the Standard Model theories.

\section*{Acknowledgments}

It is a pleasure to thank Ennio Salvioni for useful discussions, 
and Joachim Kopp and Andrea Thamm for useful comments on the manuscript. 
MJB would like to thank CERN Theory Department for warm hospitality
during part of this work.  This work has been funded by the Swiss National Science
Foundation (SNF) under contract 200021-175940 and the German Research
Foundation (DFG) under Grant Nos.\ EXC-1098, \mbox{KO~4820/1--1}, FOR~2239,
GRK~1581, and by the European Research Council (ERC) under the European Union's
Horizon 2020 research and innovation programme (grant agreement No.\ 637506,
``$\nu$Directions'').

\bibliographystyle{JHEP}
\bibliography{./paper}

\end{document}